\documentclass[usenatbib]{mn2e}
\usepackage[pdftex]{graphicx}
\newcommand{\be}{\begin{equation}}
\newcommand{\ee}{\end{equation}}
\newcommand{\bea}{\begin{eqnarray}}
\newcommand{\eea}{\end{eqnarray}}
\newcommand{\bc}{\begin{center}}
\newcommand{\ec}{\end{center}}

\def\gsim{ \lower .75ex \hbox{$\sim$} \llap{\raise .27ex \hbox{$>$}} }
\def\lsim{ \lower .75ex \hbox{$\sim$} \llap{\raise .27ex \hbox{$<$}} }

\usepackage{amsmath}
\usepackage{afterpage}

\usepackage{amssymb}

\usepackage{times}

\usepackage{paralist}

\usepackage[bookmarks,bookmarksnumbered,colorlinks=true,citecolor=blue,linkcolor=black]{hyperref}
\usepackage{color}

\usepackage{pbox}

\usepackage{booktabs}

\usepackage[flushleft]{threeparttable}

\definecolor{darkgreen}{rgb}{0.0, 0.5, 0.0}

\usepackage{grffile}

\newcommand{\Msun}{{\rm M}_{\odot}}

\newcommand{\krot}{\kappa_{\rm rot}}

\renewcommand{\thefootnote}{\fnsymbol{footnote}}

\title[Mergers and halo spin in galaxy morphology]{The role of mergers and halo spin in shaping galaxy morphology}

\author[V. Rodriguez-Gomez et al.]
{
	\parbox{18cm}{
		Vicente Rodriguez-Gomez,$^{1,2 \star}$
		Laura V. Sales,$^{3}$
		Shy Genel,$^{4,5}$
		Annalisa Pillepich,$^{1,6}$ \\
		Jolanta Zjupa,$^{7,8}$
		Dylan Nelson,$^{9}$
		Brendan Griffen,$^{10}$
		Paul Torrey,$^{10}$
		Gregory F. Snyder,$^{11}$ \\
		Mark Vogelsberger,$^{10}$
		Volker Springel,$^{7,12}$
		Chung-Pei Ma,$^{13}$
		and Lars Hernquist$^{1}$
	}
	\vspace{0.3cm} \\ 
	$^{1}$ Harvard-Smithsonian Center for Astrophysics, 60 Garden Street, Cambridge, MA 02138, USA \\
	$^{2}$ Department of Physics \& Astronomy, Johns Hopkins University, 3400 N. Charles Street, Baltimore, MD 21218, USA \\
	$^{3}$ Department of Physics \& Astronomy, University of California, Riverside, 900 University Avenue, Riverside, CA 92521, USA \\
	$^{4}$ Center for Computational Astrophysics, Flatiron Institute, 162 5th Avenue, New York, NY 10010, USA \\
	$^{5}$ Columbia Astrophysics Laboratory, Columbia University, 550 West 120th Street, New York, NY 10027, USA \\
	$^{6}$ Max-Planck-Institut f\"{u}r Astronomie, K\"{o}nigstuhl 17, D-69117 Heidelberg, Germany \\
	$^{7}$ Heidelberg Institute for Theoretical Studies, Schloss-Wolfsbrunnenweg 35, D-69118 Heidelberg, Germany \\
	$^{8}$ Institut f\"{u}r Theoretische Physik, Philosophenweg 16, D-69120 Heidelberg, Germany \\
  $^{9}$ Max-Planck-Institut f\"{u}r Astrophysik, Karl-Schwarzschild-Stra\ss{}e 1, D-85741 Garching bei M\"{u}nchen, Germany \\
	$^{10}$ Department of Physics, Kavli Institute for Astrophysics and Space Research, Massachusetts Institute of Technology, Cambridge, MA 02139, USA \\
	$^{11}$ Space Telescope Science Institute, 3700 San Martin Drive, Baltimore, MD 21218, USA \\
	$^{12}$ Zentrum f\"{u}r Astronomie der Universit\"{a}t Heidelberg, ARI, M\"onchhofstr. 12-14, D-69120 Heidelberg, Germany \\
	$^{13}$ Department of Astronomy, University of California, Berkeley, CA 94720, USA
}

\begin{document}


\maketitle
\begin{abstract}
Mergers and the spin of the dark matter halo are factors traditionally believed to determine the morphology of galaxies within a $\Lambda$CDM cosmology. We study this hypothesis by considering approximately 18,000 central galaxies at $z=0$ with stellar masses $M_{\ast} = 10^{9}$--$10^{12} \, {\rm M}_{\odot}$ selected from the Illustris cosmological hydrodynamic simulation. The fraction of accreted stars -- which measures the importance of massive, recent and dry mergers -- increases steeply with galaxy stellar mass, from less than 5 per cent in dwarfs to 80 per cent in the most massive objects, and the impact of mergers on galaxy morphology increases accordingly. For galaxies with $M_{\ast} \gtrsim 10^{11} \, {\rm M}_{\odot}$, mergers have the expected effect: if gas-poor they promote the formation of spheroidal galaxies, whereas gas-rich mergers favour the formation and survivability of massive discs. This trend, however, breaks at lower masses. For objects with $M_{\ast} \lesssim 10^{11} \, {\rm M}_{\odot}$, mergers do not seem to play any significant role in determining the morphology, with accreted stellar fractions and mean merger gas fractions that are indistinguishable between spheroidal and disc-dominated galaxies. On the other hand, halo spin correlates with morphology primarily in the {\it least} massive objects in the sample ($M_{\ast} \lesssim 10^{10} \, {\rm M}_{\odot}$), but only weakly for galaxies above that mass. Our results support a scenario where (1) mergers play a dominant role in shaping the morphology of massive galaxies, (2) halo spin is important for the morphology of dwarfs, and (3) the morphology of medium-sized galaxies -- including the Milky Way -- shows little dependence on galaxy assembly history or halo spin, at least when these two factors are considered individually.
\end{abstract}


\section{Introduction}\label{sec:intro}
\renewcommand{\thefootnote}{\fnsymbol{footnote}}
\footnotetext[1]{E-mail: vrg@jhu.edu}

\renewcommand{\thefootnote}{\arabic{footnote}}

Since \cite{Hubble1926} proposed his galaxy classification scheme, numerous studies have investigated the physical mechanisms that lead to the formation of spiral and elliptical galaxies. Given the nonlinearity of the physical processes involved, many such studies have used numerical simulation as their main tool, with \cite{Toomre1972} and \cite{White1978} being some of the first examples. To this day, one of the main questions in the field of galaxy formation remains to understand which properties of a halo (or its environment) determine the morphology of the galaxy formed at its centre \citep[e.g.][]{Parry2009, Stinson2010, Sales2012, Teklu2015, Zavala2016}.

Galactic discs are believed to form through the dissipational collapse of gas in dark matter (DM) haloes, which acquire their angular momentum through tidal torques in the early Universe \citep{Peebles1969, Doroshkevich1970}. Despite the fact that galactic outflows, misaligned gas accretion, and mergers tend to complicate the detailed conservation of specific angular momentum during the formation of a galaxy \citep[e.g.][]{VandenBosch2002, Brook2011, Zjupa2017}, it is nevertheless expected that the largest galactic discs form preferentially in haloes with higher angular momentum \citep[e.g.][]{Fall1980, Fall1983, Mo1998}. Initial attempts to verify this prediction in cosmological simulations suffered from numerical issues leading to a `catastrophic' loss of angular momentum \citep{Navarro1995}, but more recent studies have been able to verify such a trend \citep[e.g.][]{Teklu2015}.

On the other hand, elliptical galaxies are believed to be the remnants of repeated galaxy mergers \citep{Toomre1977}. This picture gained support as the hierarchical nature of structure formation \citep{White1978a} started to become recognized, although a number of objections \citep[e.g.][]{Ostriker1980} indicated that the situation was more complicated than this. Subsequent studies using idealized merger simulations have shown that the outcome of a galaxy merger is significantly affected by the masses, gas fractions, and morphologies of the progenitors, as well as by their orbital parameters \citep[e.g.][]{Barnes1992, Hernquist1992, Hernquist1993, Barnes1996, Naab2003, Cox2006}. Because of this, it is necessary to place such merger simulations in a `cosmological context' in order to evaluate the impact of galaxy mergers from a statistical perspective.

A relatively inexpensive way of doing this consists in deriving simple prescriptions from such idealized merger simulations and then combining them with semi-analytic models (SAMs) of galaxy formation \citep[e.g.][]{Khochfar2006, Parry2009, DeLucia2011, Zavala2012, Avila-Reese2014, Fontanot2015, Kannan2015} or semi-empirical halo occupation models \citep[e.g.][]{Hopkins2009a, Stewart2009a, Hopkins2010, Moster2014}. A more straightforward approach, although far more computationally expensive, is to run a hydrodynamic cosmological simulation on a large comoving volume and then examine the morphologies of the resulting galaxies directly (see \citealt{Naab2014}, section 2; \citealt{Somerville2015}, section 4.2, for reviews). With this technique, it has been found that both the numerical treatment of hydrodynamics and the feedback implementation can have a dramatic effect on galaxy morphology \citep[e.g.][]{Brook2004, Okamoto2005, Governato2007, Scannapieco2008, Scannapieco2012, Ceverino2009, Sales2010, Agertz2011, Torrey2012, Ubler2014, Snyder2015a, Genel2015, Agertz2016, Dubois2016}. Despite all the inevitable uncertainties in the modelling of feedback, the latest generation of cosmological hydrodynamic simulations \citep[e.g.][]{Vogelsberger2014, Dubois2014, Schaye2015} has been able to produce galaxy populations displaying a `morphological mix' that agrees reasonably well with observations.

The latter approach, i.e. using hydrodynamic cosmological simulations to directly investigate the origin of galaxy morphology, has presented some serious puzzles for galaxy formation. For example, by studying a sample of Milky Way-sized simulated galaxies, \cite{Sales2012} found no correlation between galaxy morphology and properties such as halo spin and merging history, challenging the `standard' model of the formation of discs and spheroids. Instead, \cite{Sales2012} proposed that a disc forms when the angular momentum of freshly accreted gas is aligned with that of earlier gas accretion. Furthermore, \cite{Zavala2016} found that galaxy morphology is correlated with the assembly history of the {\it inner} DM halo. In particular, they found that the `angular momentum loss' of a galaxy's stellar component is strongly linked to that of the inner DM halo, whereas the angular momentum history of a galaxy's cold gas is statistically correlated with that of the whole DM halo.

In this work we explore the origin of discs and spheroids using the Illustris simulation \citep{Vogelsberger2014, Vogelsberger2014a, Genel2014a, Sijacki2015}, a hydrodynamic cosmological simulation that has been shown to reproduce several galaxy observables reasonably well, including stellar angular momenta \citep{Genel2015} and quantitative optical morphologies \citep{Torrey2015, Snyder2015}, over a wide range of stellar masses. This makes the Illustris simulation a powerful tool to study the physical mechanisms that shape galaxy morphology in a cosmological context.

The results presented in this paper complement those of \cite{Genel2015}, who showed that feedback -- both from galactic winds and from active galactic nuclei -- is crucial in order to produce simulated galaxies with stellar angular momenta in reasonable agreement with observations. Here, on the other hand, we adopt the fiducial model from the Illustris simulation and focus on how the morphology of a central galaxy at $z=0$ depends on its merging history and on the spin of its host halo.

The current paper is organized as follows. In Section \ref{sec:methodology} we briefly describe the set of cosmological simulations known as the Illustris Project, along with their merger trees and other post-processed catalogues, and present the galaxy sample considered for this study. An overview of galaxy morphology in the Illustris simulation, covering a wide range of stellar masses, is presented in Section \ref{sec:measuring_morphology}. In Section \ref{sec:what_drives_galaxy_morphology} we investigate the dependence of galaxy morphology on merging history (Section \ref{subsec:merging_history}), halo spin (Section \ref{subsec:halo_spin}), and a combination of both (Section \ref{subsec:mergers_and_halo_spin}). Finally, we summarize and discuss our results in Section \ref{sec:discussion_and_conclusions}.

\section{Methodology}\label{sec:methodology}

\subsection{The Illustris and Illustris-Dark simulations}\label{subsec:illustris}

We use hydrodynamic and $N$-body simulations from the Illustris Project \citep{Vogelsberger2014, Vogelsberger2014a, Genel2014a, Sijacki2015}, a set of cosmological simulations of a periodic cube of $\sim$106.5 Mpc on a side, carried out with the moving-mesh code \textsc{arepo} \citep{Springel2010}. The main simulation considered in this study, known simply as the Illustris simulation, evolves $1820^3$ DM particles with a mass of $6.26 \times 10^6 \, \Msun$, along with approximately $1820^3$ baryonic resolution elements (stellar particles or gas cells) with an average mass of $1.26 \times 10^6 \, \Msun$. The simulation features a physical model of galaxy formation \citep{Vogelsberger2013, Torrey2014} which has been shown to reproduce various galaxy observables, including several which were not used in the calibration of the model \citep[e.g.][]{Sales2014a, Wellons2015a, Rodriguez-Gomez2015, Genel2015, Mistani2016, Cook2016}.

In addition, we consider an analogous DM-only (DMO) cosmological simulation, known hereafter as Illustris-Dark, which follows the evolution of $1820^3$ DM particles with a mass of $7.52 \times 10^6 \, \Msun$ in a cosmological box of the same size, with the same initial conditions. In Section \ref{subsec:matched_catalogues} we describe a method for matching haloes between the Illustris and Illustris-Dark simulations.

For each simulation snapshot, DM haloes are identified using the friends-of-friends (FoF) algorithm \citep{Davis1985}, which links together all particle pairs separated by less than 0.2 times the mean interparticle separation. Within each halo, gravitationally bound substructures are identified with the \textsc{subfind} algorithm \citep{Springel2001, Dolag2009a}. For the remainder of this paper, FoF groups will be referred to as \textit{haloes} and \textsc{subfind} haloes as \textit{subhaloes}. The latter category includes a central subhalo (also known as `background' subhalo) which, by definition, contains all of the gravitationally bound material in the FoF group which is \textit{not} gravitationally bound to any satellite.

We define a galaxy as being composed of the stellar and star-forming (i.e. `cold') gas components of a subhalo. Here, a gas cell is considered to be star-forming if its hydrogen particle density exceeds 0.13 cm$^{-3}$ \citep{Springel2003}. Unless otherwise noted, we measure all properties of a galaxy (e.g. stellar mass or angular momentum) throughout the entire \textsc{subfind} object -- namely, without truncating the particles found outside of a fiducial aperture such as twice the stellar half-mass radius.

Our simulations were run with a $\Lambda$ cold dark matter ($\Lambda$CDM) cosmological model with parameters $\Omega_{\rm m} = 0.2726$, $\Omega_{\Lambda} = 0.7274$, $\Omega_{\rm b} = 0.0456$, $\sigma_8 = 0.809$, $n_{\rm s} = 0.963$, and $h = 0.704$, consistent with the nine-year \textit{Wilkinson Microwave Anisotropy Probe} measurements \citep{Hinshaw2013}.

\subsection{Merger trees and stellar assembly catalogues}\label{subsec:merger_trees}

We use the \textsc{sublink} algorithm \citep{Rodriguez-Gomez2015} to connect galaxies across the 136 snapshots produced by our simulations, resulting in data structures known as a merger trees. First, for each galaxy from a given snapshot, a \textit{descendant} is identified in the next snapshot by matching the stellar particles and star-forming gas cells\footnote{The `default' version of the \textsc{sublink} algorithm, included in the Illustris public data release \citep{Nelson2015}, tracks only DM particles. However, the `baryonic' version of merger trees is better suited for this study, as well as for most other galaxy formation applications \citep[see][for details]{Rodriguez-Gomez2015}.} in a weighted fashion, putting more weight on elements which are more tightly bound. In some cases, a small galaxy is allowed to `skip' a snapshot when finding a descendant, in order to avoid losing track of it while it is passing through a larger, denser structure. After assigning all the descendants, the \textit{main progenitor} of each galaxy is defined as the one with the `most massive history' behind it \citep{DeLucia2007}. This information is enough to construct the merger trees, which are stored in a `depth-first' fashion \citep{Lemson2006} in order to allow fast retrieval of the full evolutionary histories.

These merger trees have been used to determine the `accretion origin' of every stellar particle in the Illustris simulation, which results in our so-called stellar assembly catalogues \citep{Rodriguez-Gomez2016}. These datasets provide a precise measurement of the \textit{ex situ} stellar mass fraction of each galaxy, denoted by $f_{\rm acc}$, which measures the fraction of a galaxy's stellar mass contributed by stars that formed in other galaxies and which were subsequently accreted.

\subsection{Matched halo catalogues}\label{subsec:matched_catalogues}

In order to isolate (or exclude) the effects of baryons on DM, it is often useful to match (sub)haloes from a hydrodynamic simulation to their counterparts in an analogous $N$-body simulation. We carry out such a task for the Illustris and Illustris-Dark simulations in the following way. For each subhalo in Illustris, we define `matched' subhalo candidates in Illustris-Dark as those which have at least one DM particle in common. For each candidate, we evaluate the merit function
\begin{equation}
\chi = \sum_j \left({\cal R}_{j, {\rm FP}}^{-\alpha} + {\cal R}_{j, {\rm DMO}}^{-\alpha}\right),
\label{eq:merit_function}
\end{equation}
where ${\cal R}_{j, {\rm FP}}$ and ${\cal R}_{j, {\rm DMO}}$ are the binding energy ranks of the $j$-th DM particles in the `full-physics' (FP) and DMO runs, respectively, and the sum is carried out over all common DM particles. The exponent is chosen to be $\alpha = 0$ for subhaloes which are centrals in the DMO run, and $\alpha = 1$ for satellites. This choice maximizes the probability that a central is matched to a central, and a satellite to a satellite.\footnote{Equation (\ref{eq:merit_function}) is very similar to the merit function used in \cite{Rodriguez-Gomez2015} to construct merger trees, except for the symmetry with respect to the two sets of subhaloes being considered (in this case those from the FP and DMO simulations) and the flexible choice of the $\alpha$ exponent.} As a final step, the same matching procedure is applied in reverse, i.e. by matching subhaloes from the DMO simulation to their counterparts in the FP run. Only those subhaloes with a bijective match are ultimately stored in the catalogue.

At $z=0$, we find that 19,370 out of the 19,375 central galaxies in Illustris with $M_{\ast} > 10^{9} \, \Msun$ have a bijective match. However, 2--3 per cent of these objects are classified as satellites in the DMO simulation and are therefore excluded in some parts of our analysis, as explained below.

\subsection{The galaxy sample}\label{subsec:galaxy_sample}

We consider all central galaxies at $z=0$ with stellar masses $M_{\ast} > 10^{9} \Msun$. After removing a small spurious component, as explained below, our sample ultimately consists of 18,076 (17,650 in some calculations) central galaxies at $z=0$. We do not consider satellite galaxies because a connection with the spin parameter of their host haloes would be difficult to interpret.

In order to account for two common problems that arise when connecting simulated galaxies across cosmic time, we apply the following small corrections to our galaxy sample. Firstly, we remove `orphan' galaxies by making sure that all galaxies in the sample can be tracked back in time to at least $z=2$. This removes 16 out of the 19,375 central galaxies originally included in the sample. Secondly, as a consequence of the halo identification procedure, a central and a satellite can sometimes `swap' identities during close interactions \citep{Srisawat2013a, Avila2014}. In order to minimize the effects from such spurious identifications, we only consider central galaxies which have never been classified as satellites during ten or more consecutive snapshots. This restriction removes a further 1,283 galaxies from our sample, or nearly 7 per cent, leaving a total of 18,076 central galaxies at $z=0$. We find that these corrections have a negligible effect on the statistical trends shown in this paper.

Finally, when considering a DMO-matched quantity (namely, the spin of the DM halo), 426 central galaxies are further removed because their DMO counterparts are classified as satellites (see Section \ref{subsec:matched_catalogues}), reducing our sample to 17,650 central galaxies in these cases.

\section{Galaxy morphology revisited}\label{sec:measuring_morphology}

Galaxy morphology is known to correlate with various galactic properties such as the amount of rotation, star formation, gas content, and the spatial distribution of light \citep[e.g.][]{Kauffmann2003}. In \cite{Snyder2015}, non-parametric optical morphologies were calculated for thousands of Illustris galaxies, and they were found to be in good agreement with their observational counterpart \citep{Lotz2008a}. Furthermore, \cite{Genel2015} found that the specific stellar angular momenta of Illustris galaxies are consistent with observational trends \citep{Romanowsky2012, Fall2013}. All of this provided evidence that the Illustris simulation can be a powerful tool to study galaxy morphology in a cosmological context. In the current work, however, we will exclusively consider \textit{kinematic} morphologies rather than optical ones, as the former seem somewhat easier to interpret from a theoretical perspective. A detailed comparison between different measures of galaxy morphology in hydrodynamic simulations is deferred to future work.

\subsection{The amount of rotational support}\label{subsec:morphology_definitions}

\begin{figure*}
  \centering
	\includegraphics[width=17.5cm]{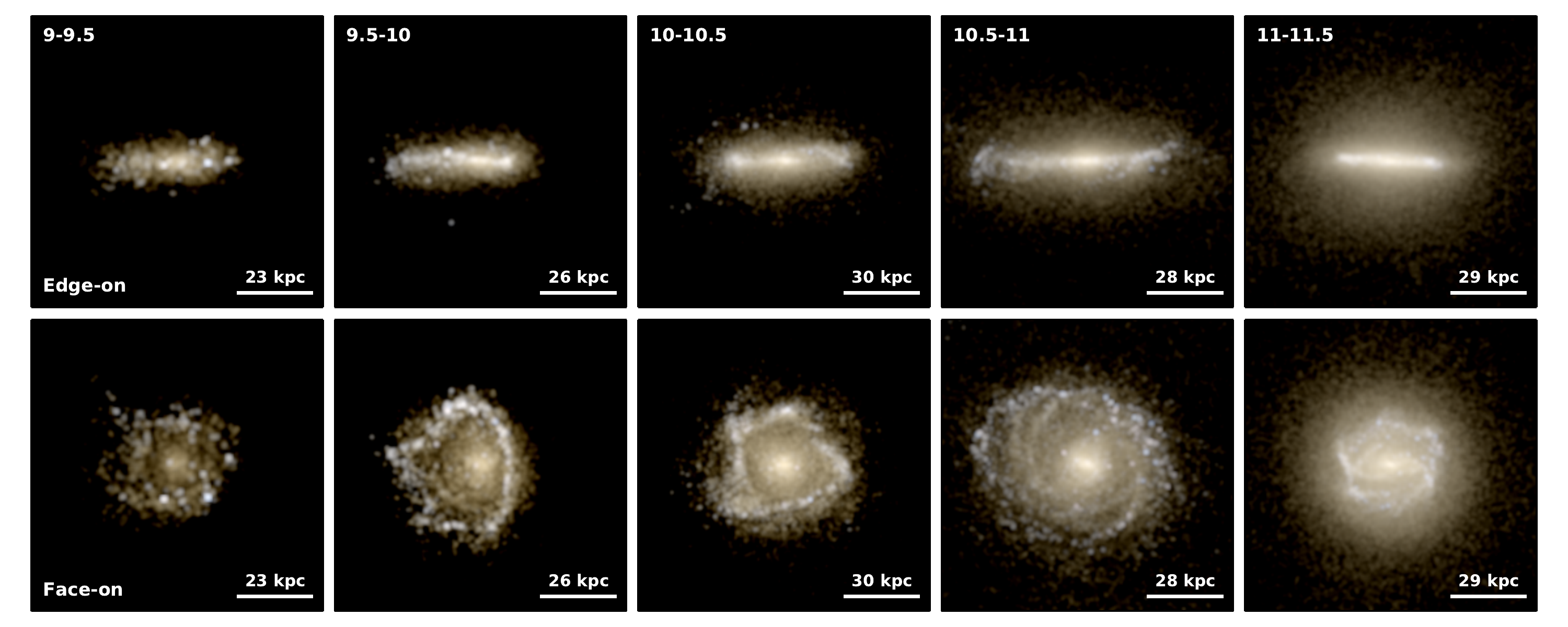}
	\caption{Disc-like galaxies of different masses, selected as those that satisfy $\kappa_{\rm rot} > 0.6$. Each column corresponds to a randomly selected disc-like galaxy at $z=0$ within the logarithmic stellar mass bin indicated in the upper left corner (in units of $\Msun$). The upper and lower panels show edge-on and face-on projections of the same galaxy, respectively. Each image shows the stellar light distribution (a composite of SDSS $g$,$r$,$i$ bands) projected onto a square with 8 stellar half-mass radii on a side. The scale in the lower-right corner of each panel corresponds to 2 stellar half-mass radii.}
	\label{fig:disks}
\end{figure*}

\begin{figure*}
  \centering
	\includegraphics[width=17.5cm]{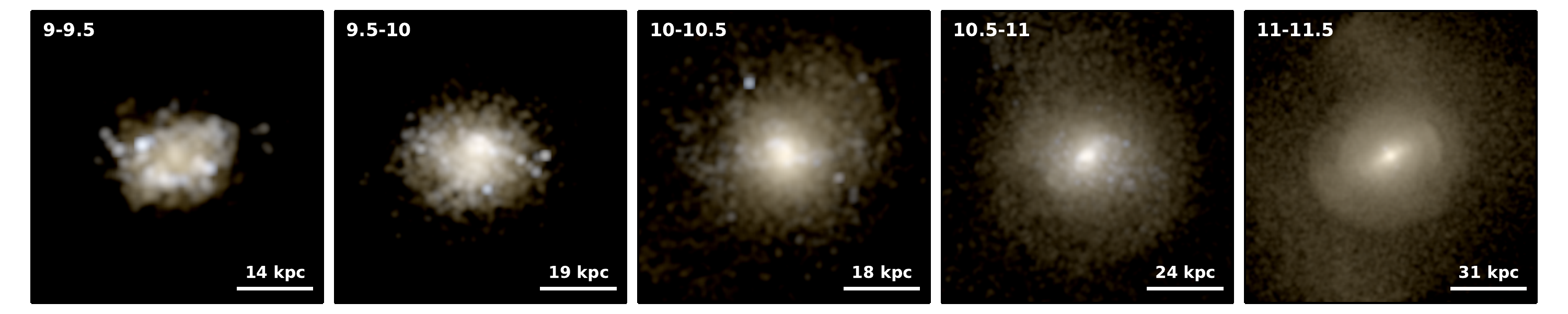}
	\caption{Same as Fig. \ref{fig:disks}, but showing spheroidal galaxies (face-on projection only) in the same stellar mass bins at $z=0$. This figure shows that increasingly massive spheroidal galaxies become more concentrated toward their centres.}
	\label{fig:spheroids}
\end{figure*}

Throughout this work we quantify the morphology of a galaxy using the $\kappa_{\rm rot}$ parameter, defined as the fraction of the total kinetic energy contributed by the azimuthal component of the stellar velocities, where the $z$-axis coincides with the total angular momentum of the galactic stellar component. More explicitly, this parameter was introduced in \cite{Sales2012} as

  \begin{equation}
    \kappa_{\rm rot} = \frac{K_{\rm rot}}{K} = \frac{1}{K}\sum_{i}\frac{1}{2} m_{i}\left(\frac{j_{z,i}}{R_{i}}\right)^{2},
    \label{eq:kappa}
  \end{equation}
where $K$ is the total kinetic energy of the stellar component, $m_i$ represents the mass of a particle, $j_{z,i}$ is the $z$-component of the specific angular momentum, $R_i$ is the projected radius, and the sum is carried out over all stellar particles in the galaxy.\footnote{In the original definition from \cite{Sales2012}, only stars within twice the stellar half-mass radius were considered. We find that both definitions yield qualitatively similar results.} Following \cite{Genel2015}, the calculation frame is centred at the position of the most bound stellar particle, while the velocity of the frame coincides with that of the stellar centre of mass. As discussed in Appendix \ref{app:kappa_expanded}, we find that $\krot$ is a good proxy for the amount of rotational support in a galaxy.

We note that $\kappa_{\rm rot}$ is strongly correlated with other measures of \textit{kinematic} morphology \citep[e.g.][figure 1]{Sales2012}. In fact, throughout this paper $\kappa_{\rm rot}$ will represent a whole class of kinematic morphological parameters, most of which are based on the distribution of the `orbital circularity' parameter $\epsilon$. This parameter is usually defined as $\epsilon = j_{z} / j(E)$, where $j(E)$ is the maximum specific angular momentum possible for a star with specific binding energy $E$ \citep{Abadi2003}, or as $\epsilon_{V} = j_{z} / r v_c(r)$, where $v_c(r) = \sqrt{G M (< r) / r}$ is the circular velocity at the distance $r$ \citep{Scannapieco2009, Scannapieco2012}. Some examples of such circularity-based morphological parameters include the disc-to-total ratio, defined as the fraction of stars with sufficiently circular orbits, typically ${\rm D/T} = f(\epsilon > 0.7) \approx f(\epsilon_{V} > 0.8)$ \citep{Aumer2013, Marinacci2013}, or the bulge-to-total fraction, usually defined as ${\rm B/T} = 2 \times f(\epsilon < 0)$ \citep[but see][for an improved B/T measurement]{Martig2012, Zavala2016}. We have found that $\kappa_{\rm rot}$ is strongly correlated with any of these alternative measures of kinematic morphology, despite the fact that $\kappa_{\rm rot}$ cannot distinguish the sense of rotation of a particle about the $z$-axis, i.e. corotating or counter-rotating (see Appendix \ref{app:kappa_expanded}). Furthermore, $\kappa_{\rm rot}$ has the advantage of being independent from particular choices of $\epsilon$ values that are considered to constitute `circular' orbits, or from the assumption that the bulge component is symmetric with respect to the $\epsilon$-parameter.

As expected, the $\kappa_{\rm rot}$ parameter is also connected to the specific angular momentum of the stars in a galaxy. Such a correlation between galaxy morphology and specific angular momentum is clearly seen in observations \citep[e.g.][]{Fall1983, Romanowsky2012, Obreschkow2014}, as well as in hydrodynamic simulations of a $\Lambda$CDM cosmology \citep[e.g.][]{Genel2015, Teklu2015, Sokolowska2016, Lagos2017}. We caution, however, that most of the stellar angular momentum in Illustris galaxies with $M_{\ast} \gtrsim 10^{11} \, \Msun$ is found at very large radii, which means that the \textit{total} stellar angular momentum does not directly measure the degree of rotational support in the central, most luminous regions of such galaxies. The relationship between $\kappa_{\rm rot}$ and angular momentum is discussed in more detail in Appendix \ref{app:angular_momentum}.

\subsection{Visual impression}\label{subsec:morphology_visual_impression}

Fig. \ref{fig:disks} shows images of randomly selected disc-like galaxies at $z=0$ in different stellar mass bins, ranging from $\log_{10}(M_{\ast}/{\rm M}_{\odot}) = 9$--$9.5$ (first column) to $\log_{10}(M_{\ast}/{\rm M}_{\odot}) = 11$--$11.5$ (last column). The upper and lower panels show edge-on and face-on projections of the same galaxies, respectively. These disc-like galaxies have been selected as those with $\kappa_{\rm rot} > 0.6$ (typically $\kappa_{\rm rot} \approx 0.6$--$0.65$), which constitutes the upper tail of the $\kappa_{\rm rot}$ distribution, as we will see in Section \ref{subsec:morphology_distribution}.

Each image is a composite of Sloan Digital Sky Survey (SDSS; \citealt{York2000}) $g,r,i$ filters, calculated using stellar population synthesis models \citep[e.g.][]{Bruzual2003}. The scaling of the images is logarithmic (or, equivalently, linear on a magnitude scale) as described in \cite{Lupton2004}, and the light from each stellar particle has been smoothed using a Gaussian kernel with a standard deviation equal to the gravitational softening length of baryonic resolution elements (0.7 kpc at $z=0$). In principle, these images could also be generated with adaptive softening as described in \cite{Torrey2015}.

Similarly, Fig. \ref{fig:spheroids} shows projected stellar density maps of randomly selected spheroidal galaxies at $z=0$ in the same stellar mass bins. These spheroidal systems have been selected as those satisfying $\kappa_{\rm rot} < 0.3$ (typically $\kappa_{\rm rot} \approx 0.25$--$0.3$). We can see that, as their stellar mass increases, spheroidal systems become more concentrated toward their centres, and some of them even display signs of shells and tidal streams.

Figs. \ref{fig:disks} and \ref{fig:spheroids} demonstrate that discs and spheroids arise naturally in the Illustris simulation over a wide range of stellar masses. We note, however, that these galaxy images merely represent the tails of the $\kappa_{\rm rot}$ spectrum. In general, we find a smooth transition from disc-like to spheroidal systems.

\subsection{Distribution of galaxy morphologies}\label{subsec:morphology_distribution}

\begin{figure}
  \centering
	\includegraphics[width=8cm]{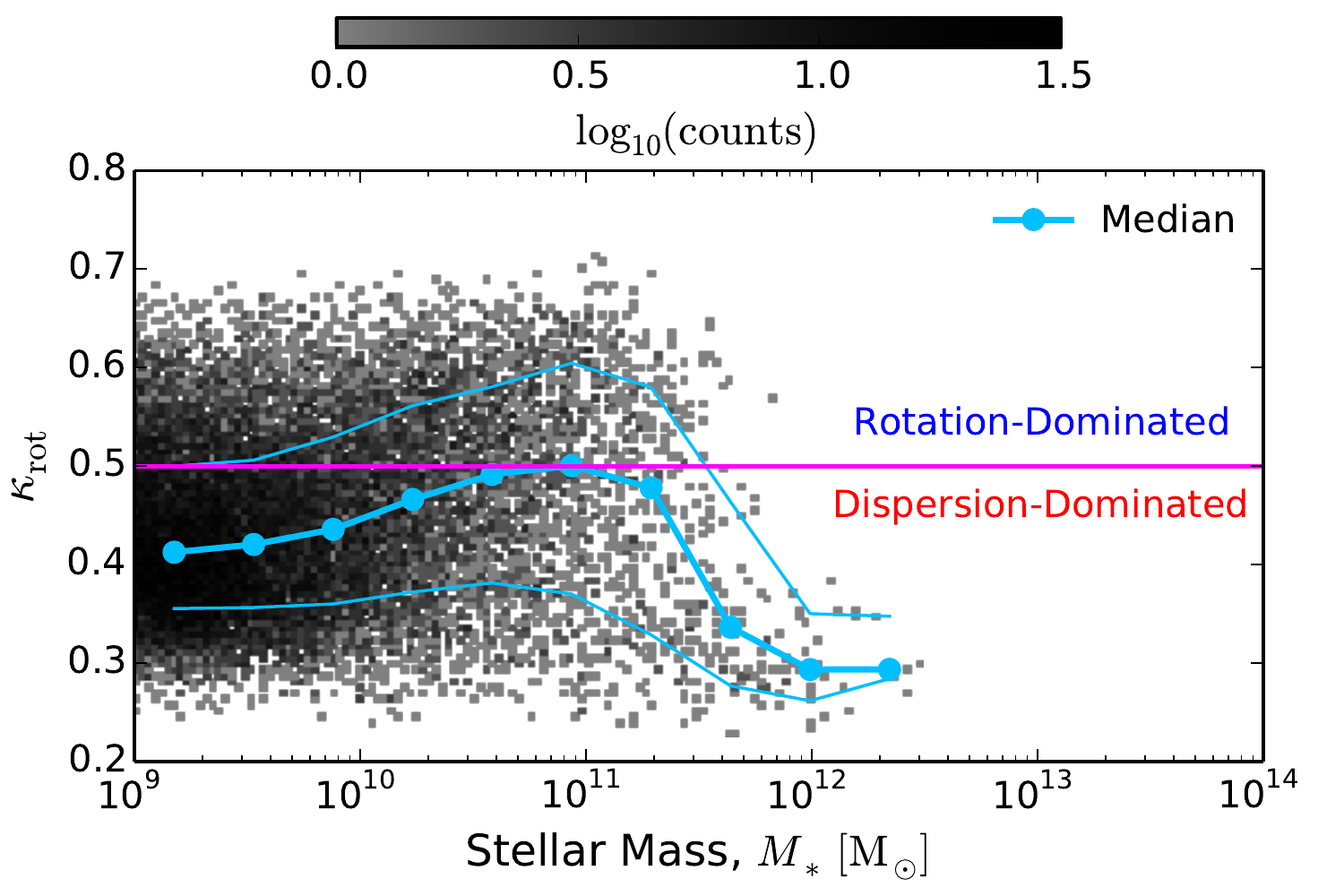}
	\caption{The $\kappa_{\rm rot}$ morphological parameter as a function of stellar mass, calculated for central galaxies at $z=0$. In this and similar plots, a two-dimensional histogram with very small bins serves the purpose of a scatter plot. The (small) bins are coloured so that a darker shade of grey corresponds to a larger number of galaxies per bin. The thick cyan line shows the median trend of $\kappa_{\rm rot}$ as a function of stellar mass, while the thin cyan lines indicate the 16th to 84th percentile range, or $1\sigma$. The horizontal magenta line is located at $\kappa_{\rm rot} = 0.5$, separating galaxies into rotation-dominated and dispersion-dominated morphological types, as indicated by the blue and red text labels.}
	\label{fig:morphology_vs_mass}
\end{figure}

In Fig. \ref{fig:morphology_vs_mass}, the two-dimensional histogram (effectively a scatter plot, given the small bin size) shows the distribution of galaxies with respect to $\kappa_{\rm rot}$ and stellar mass, with darker shades of grey indicating a larger number of galaxies per bin. The use of a two-dimensional histogram to represent a scatter plot reduces the loss of information that is inevitable in the most densely populated regions of a `normal' scatter plot. The thick cyan line shows the median value of $\kappa_{\rm rot}$ as a function of stellar mass, while the thin cyan lines indicate the 16th to 84th percentile range, or $1\sigma$. The horizontal magenta line is located at $\kappa_{\rm rot} = 0.5$. Hereafter, we refer to galaxies with $\kappa_{\rm rot} \geq 0.5$ as rotation-dominated and to those with $\kappa_{\rm rot} < 0.5$ as dispersion-dominated. Although any $\kappa_{\rm rot}$ value between 0 and 1 is possible in theory, as discussed in Appendix \ref{app:kappa_expanded}, in practice $\kappa_{\rm rot}$ takes values between $\kappa_{\rm rot} \approx 0.25$ and $\kappa_{\rm rot} \approx 0.7$ for Illustris galaxies.

We note that the distribution of kinematic morphologies shown in Fig. \ref{fig:morphology_vs_mass} is remarkably similar to the one found by \cite{Dubois2016} using the Horizon-AGN simulation \citep{Dubois2014}, despite the different (but also kinematic) definition of galaxy morphology used in their study, and the obviously different numerical methods used to model the hydrodynamics and other physical processes. In particular, the simulated galaxies from \cite{Dubois2016} as well as those from the current study are more likely to be disc-like at $z=0$ if their stellar masses are around $M_{\ast} \approx 10^{10.5}$--$10^{11} \Msun$.

\section{What drives galaxy morphology?}\label{sec:what_drives_galaxy_morphology}

In this section we examine the connection between galaxy morphology, merging history, and halo spin across a wide range of stellar masses.

\subsection{Merging history}\label{subsec:merging_history}

Idealized merger simulations have shown that mergers can play an important role in determining the morphology of a galaxy \citep[e.g.][]{Toomre1977, White1978, Barnes1996}. However, the `cosmologically averaged' effect of mergers on galaxy morphology, as well as the relative impact on morphology produced by major versus minor mergers, or gas-rich versus dry mergers, are still the subject of significant discussion (e.g. \citealt{Khochfar2003, Naab2006a, Hopkins2009b, Hopkins2010, Fiacconi2014}; see \citealt{Naab2014}, for a review). Here we explore these topics using a large, cosmologically representative sample of simulated galaxies at $z=0$ (defined in Section \ref{subsec:galaxy_sample}), with the help of merger trees and other useful data products described in Section \ref{sec:methodology}.

\subsubsection{The \textit{ex situ} stellar mass fraction}\label{subsec:ex_situ_fraction}

The main parameter that we use to quantify the importance of merging history is the \textit{ex situ} stellar mass fraction, denoted by $f_{\rm acc}$ and defined as the fraction of a galaxy's stellar mass contributed by stars that formed in other galaxies and which were subsequently accreted. This quantity has already been measured to great precision in the Illustris simulation \citep{Rodriguez-Gomez2016}, where it was found to strongly correlate with stellar mass and other galaxy properties.

It is important to note that $f_{\rm acc}$ does not measure the merging history \textit{per se}, but instead quantifies the relative importance of dry merging with respect to dissipative processes such as \textit{in situ} star formation \citep[e.g.][]{Oser2010}. Because of star formation quenching in massive galaxies, the \textit{ex situ} fraction $f_{\rm acc}$ displays a much stronger variation with stellar mass than the merger rate itself, which shows a positive but relatively mild dependence on stellar mass \citep{Rodriguez-Gomez2015}.

\begin{figure}
  \centerline{\hbox{
	\includegraphics[width=8cm]{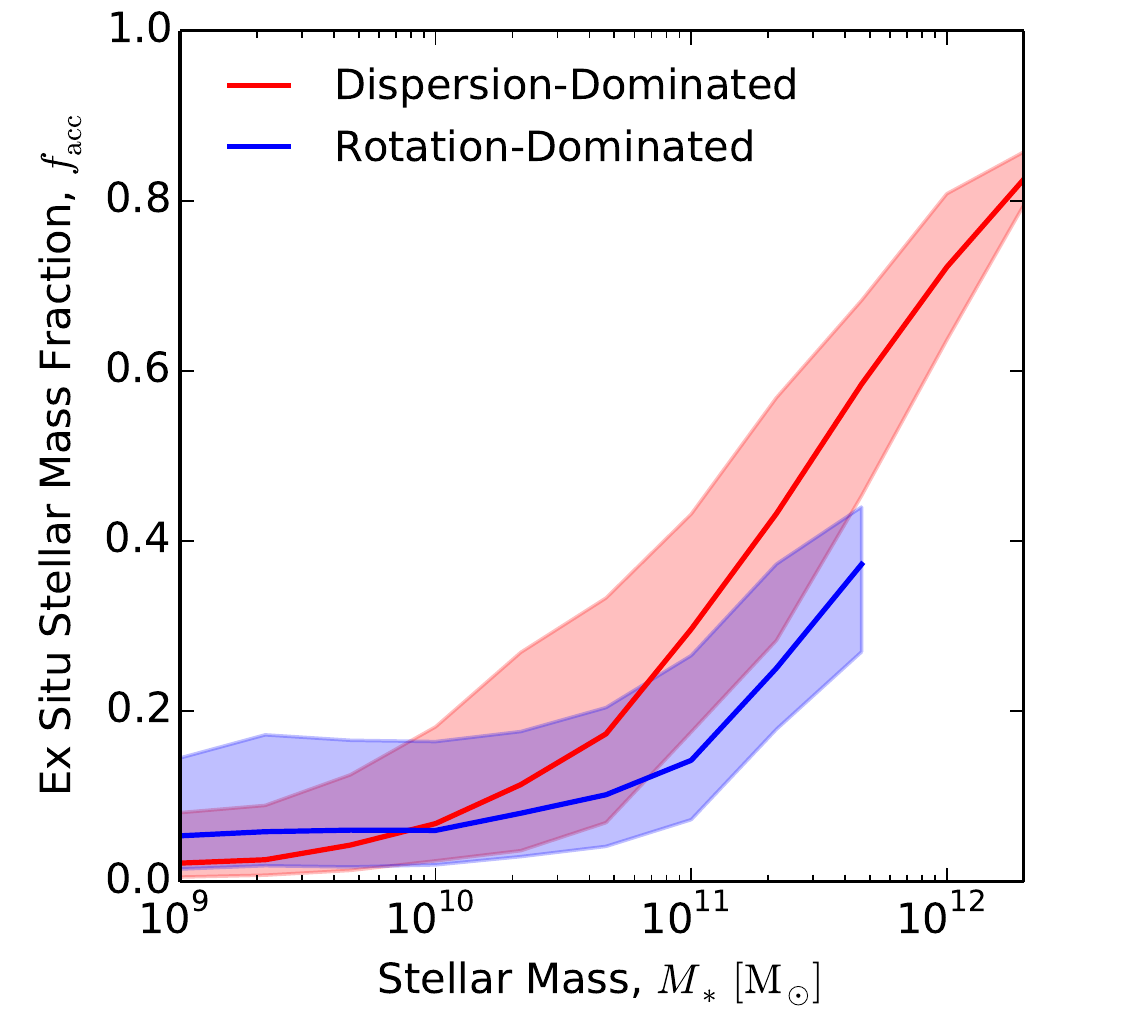}
  }}
	\caption{The median \textit{ex situ} stellar mass fraction $f_{\rm acc}$ as a function of stellar mass $M_{\ast}$, shown for central galaxies at $z=0$. The galaxies have been separated into two morphological types according to their $\kappa_{\rm rot}$ values. The shaded regions indicate the 16th to 84th percentile ranges.}
	\label{fig:f_acc_vs_mstar}
\end{figure}

In Fig. \ref{fig:f_acc_vs_mstar} we show the median \textit{ex situ} stellar mass fraction as a function of stellar mass $M_{\ast}$, distinguishing between dispersion-dominated ($\kappa_{\rm rot} < 0.5$, red) and rotation-dominated ($\kappa_{\rm rot} \geq 0.5$, blue) central galaxies at $z=0$. The shaded regions indicate the 16th to 84th percentile range, or 1$\sigma$, at a fixed stellar mass. Similar plots have already been shown in \cite{Rodriguez-Gomez2016}, except that we now separate galaxies by adopting a fixed threshold at $\kappa_{\rm rot} = 0.5$, instead of using the median value of $\kappa_{\rm rot}$ at a fixed stellar mass.

Fig. \ref{fig:f_acc_vs_mstar} shows that dry mergers play an increasingly important role at higher masses, where dispersion-dominated galaxies have larger \textit{ex situ} stellar mass fractions than rotation-dominated galaxies of similar masses, as expected within the framework of $\Lambda$CDM cosmology and the merger hypothesis. A less expected result is that the correlation between $\kappa_{\rm rot}$ and $f_{\rm acc}$ becomes very weak at $M_{\ast} \lesssim 10^{11} \Msun$. Interestingly, this is consistent with \cite{Sales2012}, who found a null correlation between $\kappa_{\rm rot}$ and $f_{\rm acc}$ for galaxies in a similar mass range using the \textsc{gimic} simulation \citep{Crain2009}.

In Fig. \ref{fig:kappa_vs_f_acc} we further explore the statistical relationship between galaxy morphology and the \textit{ex situ} stellar mass fraction. The scatter plots (or, more precisely, two-dimensional histograms with very small bins) show the galaxy distribution with respect to $\kappa_{\rm rot}$ and $f_{\rm acc}$ in different stellar mass bins. The upper panels show the marginal distribution with respect to $\kappa_{\rm rot}$, separating between dispersion-dominated ($\kappa_{\rm rot} < 0.5$, red) and rotation-dominated ($\kappa_{\rm rot} \geq 0.5$, blue) systems. This figure shows even more explicitly that $\kappa_{\rm rot}$ and $f_{\rm acc}$ are correlated at high masses, and that the trend fades out at lower masses.

\subsubsection{Other merger statistics}\label{subsec:merger_statistics}

\begin{figure*}
  \centering
	\includegraphics[width=17.5cm]{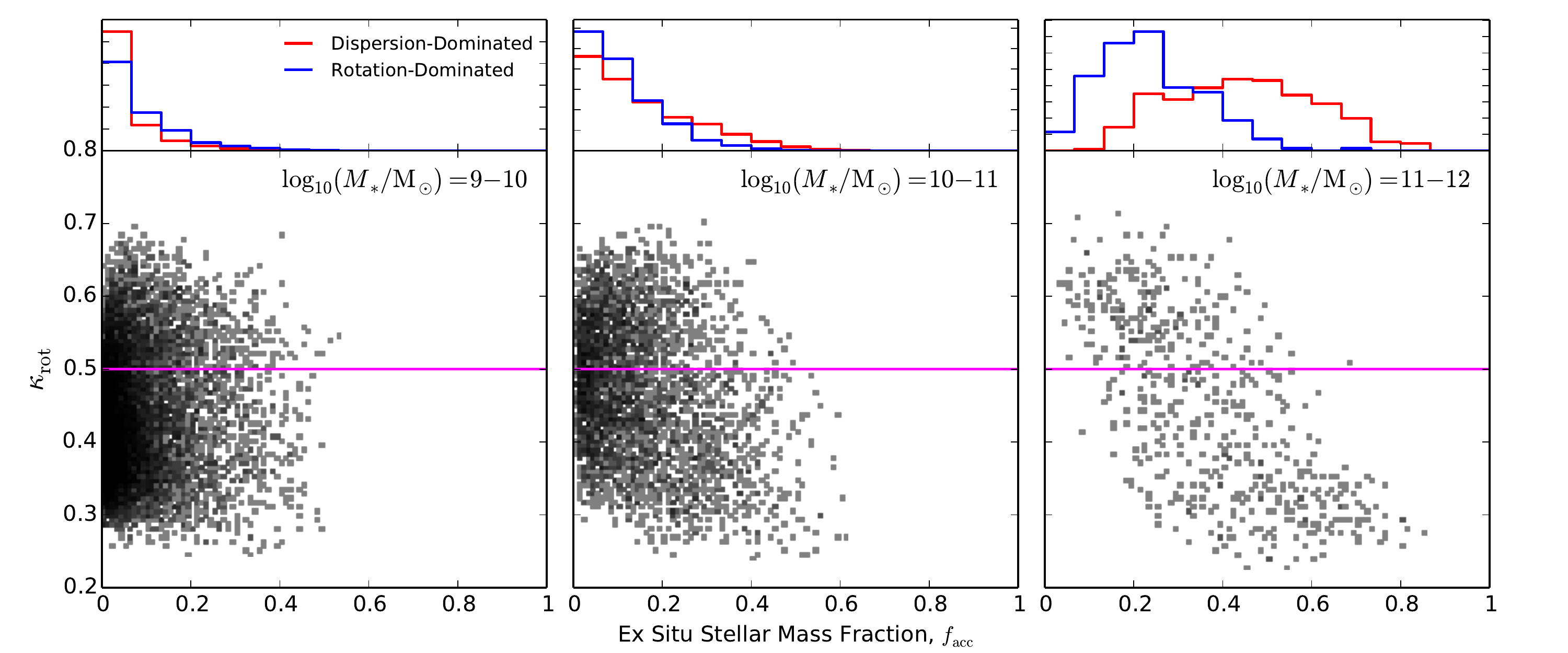}
	\caption{Correlation between galaxy morphology and the \textit{ex situ} stellar mass fraction $f_{\rm acc}$ in different stellar mass bins at $z=0$. The scatter plots (or two-dimensional histograms with very small bins, using the colour scale from Fig. \ref{fig:morphology_vs_mass}) in the lower panels show the overall galaxy distribution, while the histograms in the upper panels show the corresponding marginal distributions for early- and late-type galaxies, which have been classified according to the horizontal magenta line in the lower panels. This figure demonstrates that galaxy morphology is clearly correlated with $f_{\rm acc}$ in the most massive bin, but less so in the case of smaller galaxies.}
	\label{fig:kappa_vs_f_acc}
\end{figure*}

\begin{figure*}
  \centering
	\includegraphics[width=17.5cm]{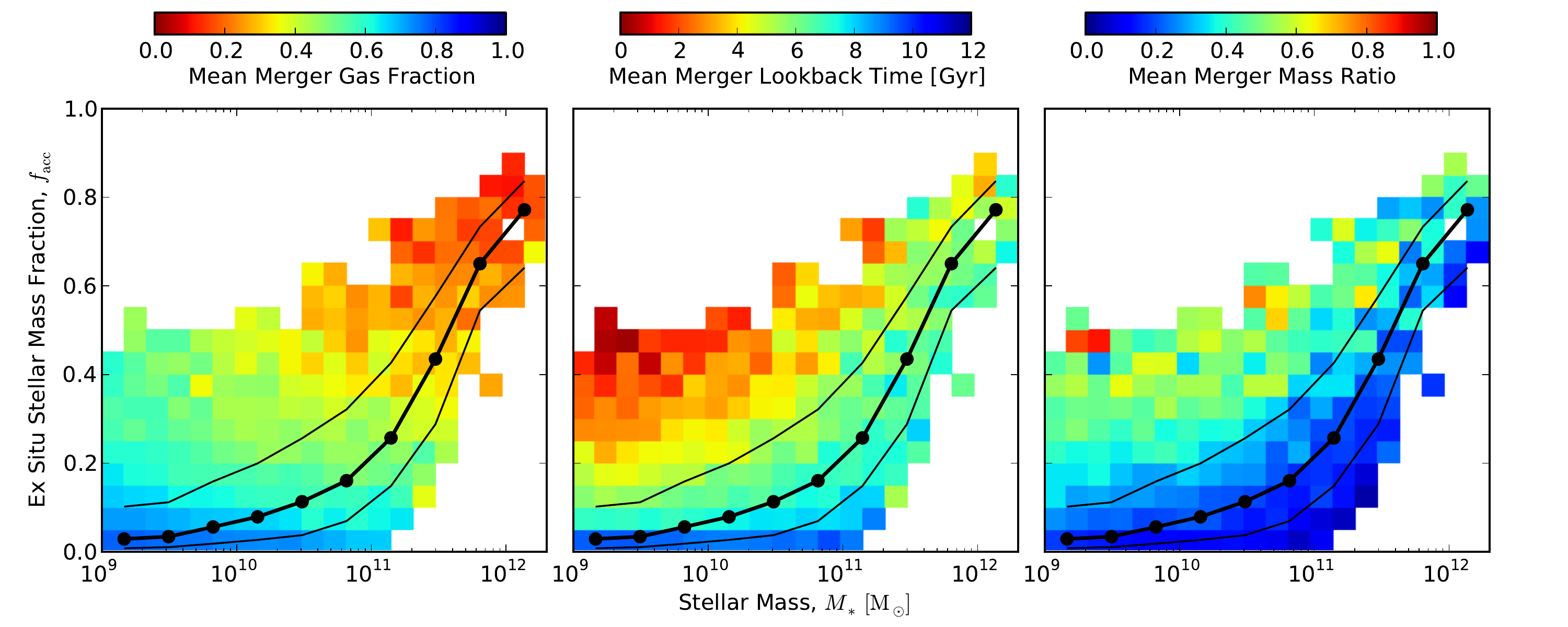}
	\caption{The \textit{ex situ} stellar mass fraction $f_{\rm acc}$ as a function of stellar mass $M_{\ast}$ for central galaxies at $z=0$. The two-dimensional histograms are coloured according to the median value of different merger statistics (different panels) of the galaxies that fall into each bin. Such merger statistics are the mean (mass-weighted) cold gas fraction of the secondary progenitors from all mergers (left), the mean (mass-weighted) lookback time of all mergers (centre), and the mean (mass-weighted) stellar mass ratio of all mergers (right). In each panel, the thick black line shows the median trend of $f_{\rm acc}$ as a function of stellar mass, while the thin lines show the 16th to 84th percentile range. At any fixed stellar mass, there is a positive correlation between $f_{\rm acc}$ and each of the merger statistics shown in the different panels. This demonstrates that, to a first approximation, $f_{\rm acc}$ is a good proxy for \textit{massive}, \textit{recent}, and \textit{dry} merging history. Also note that dry mergers become ubiquitous at $M_{\ast} > 10^{11} \Msun$.}
	\label{fig:merging_history}
\end{figure*}

\begin{figure*}
  \centering
	\includegraphics[width=17.5cm]{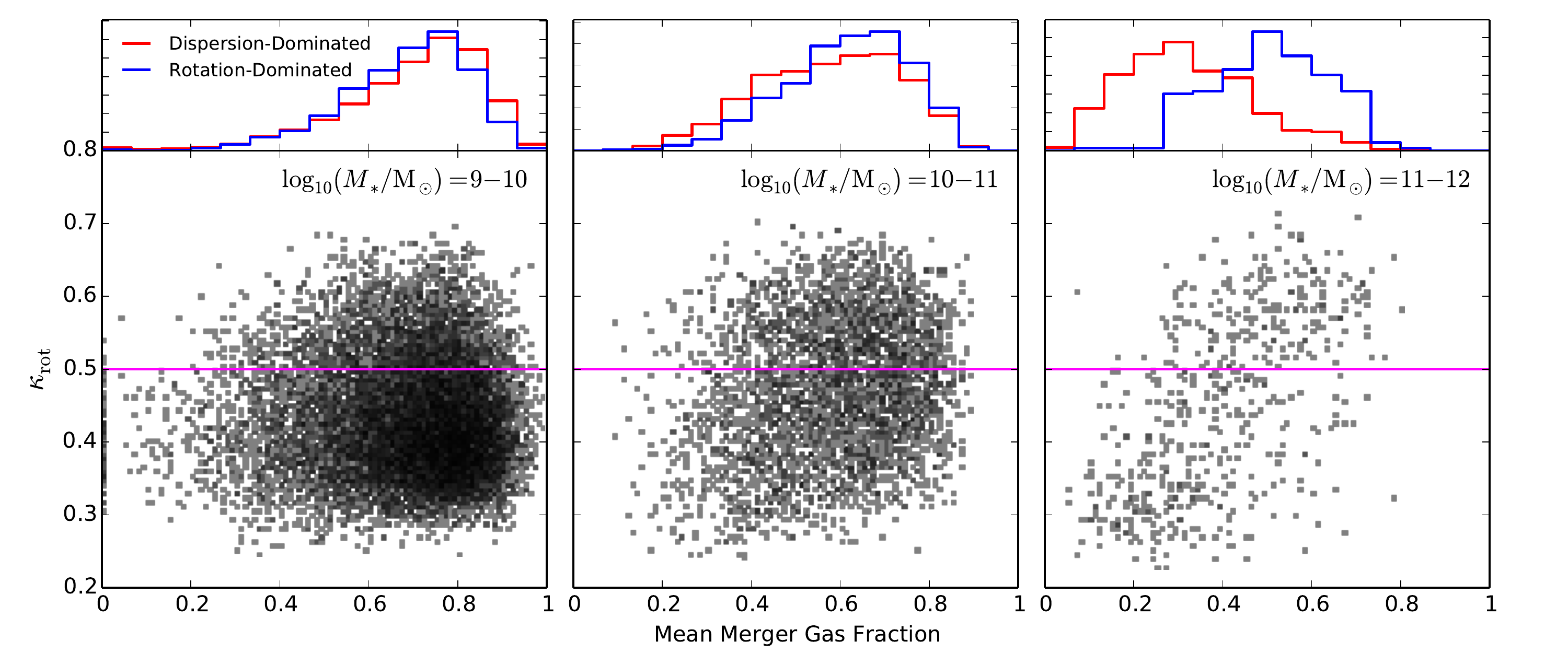}
	\caption{Correlation between galaxy morphology and the cold gas fraction of the baryonic material brought in by mergers, shown for different stellar mass bins at $z=0$. The scatter plots in the lower panels show the overall galaxy distribution, while the histograms in the upper panels show the corresponding marginal distributions of early- and late-type galaxies. In agreement with results from idealized merger simulations, dry mergers tend to produce dispersion-dominated systems, while remnants from gas-rich mergers can possess a significant disc component. This trend is only significant in the right-hand panel, which corresponds to the mass range in which mergers have a definite impact on galaxy morphology (compare with Fig. \ref{fig:kappa_vs_f_acc}).}
	\label{fig:kappa_vs_merger_gas_fraction}
\end{figure*}

In addition to the \textit{ex situ} stellar mass fraction, which we consider to be a `first order' indicator of a galaxy's merging history, we also define and explore the following merger statistics:

\begin{itemize}

\item \textit{Mean merger gas fraction:} the mean `cold' (i.e. star-forming) gas fraction of all the objects that have merged with the galaxy in question, weighted by their maximum stellar masses. The cold gas fraction of each secondary progenitor is measured at the time when it reached its maximum stellar mass. \\

\item \textit{Mean merger lookback time:} the mean lookback time of all the mergers that a galaxy has undergone, weighted by the maximum stellar mass of the secondary progenitors. Each merger is assumed to have taken place at the time when the two branches of the merger tree merged with each other. \\

\item \textit{Mean merger mass ratio:} the mean stellar mass ratio of all the mergers that a galaxy has undergone, weighted by the maximum stellar mass of the secondary progenitors. Following \cite{Rodriguez-Gomez2015}, each merger mass ratio is measured at the time when the secondary progenitor reached its maximum stellar mass.

\end{itemize}

In Fig. \ref{fig:merging_history}, the two-dimensional histograms in the different panels are coloured according to the median value of the three merger statistics described above, with the horizontal and vertical axes corresponding to $M_{\ast}$ and $f_{\rm acc}$, respectively. For reference, the thick and thin black lines show the median and 1$\sigma$ scatter of $f_{\rm acc}$ as a function of stellar mass.

Apart from the evident trend between $f_{\rm acc}$ and stellar mass, which has already been discussed in \cite{Rodriguez-Gomez2016}, Fig. \ref{fig:merging_history} shows that -- at any fixed stellar mass -- $f_{\rm acc}$ is negatively correlated with the mean merger gas fraction and the mean lookback time (i.e. gas-rich mergers, as well as those which happened a longer time ago, tend to be associated with lower \textit{ex situ} fractions) and positively correlated with the mean merger mass ratio (i.e. major mergers lead to higher $f_{\rm acc}$ values). Therefore, as a first approximation, we consider $f_{\rm acc}$ to be a good measure of the importance of \textit{dry}, \textit{recent}, and \textit{massive} mergers (relative to \textit{in situ} star formation). This justifies the `encapsulation' of merging history -- at least the kind of merging history that leads to spheroid formation -- into a single number, i.e. the \textit{ex situ} stellar mass fraction $f_{\rm acc}$.

Besides the correlation between $f_{\rm acc}$ and each of the three additional merger statistics, a closer inspection of Fig. \ref{fig:merging_history} reveals other interesting trends. First, dry mergers are prevalent for all galaxies with $M_{\ast} > 10^{11} \Msun$. An implication of this is that two galaxies with the same $f_{\rm acc}$ value but with different stellar masses need not have `equivalent' merging histories. For example, Fig. \ref{fig:merging_history} shows that galaxies with $f_{\rm acc} \approx 0.4$ at the low-mass end of our sample are extreme outliers which appear to have undergone a single, recent, gas-rich major merger (note that these galaxies also have mean merger mass ratios around 0.4). On the other hand, the typical merging history of a very massive galaxy with $f_{\rm acc} \approx 0.4$ consists of numerous dry, minor mergers. Therefore, although we often use $f_{\rm acc}$ as a simplified measure of the importance (relative to \textit{in situ} star formation) of a galaxy's merging history, a fixed $f_{\rm acc}$ value can have very different implications for galaxies of different masses.

Another observation about Fig. \ref{fig:merging_history} worth mentioning is that, for the `bulk' of the galaxy population (say, the galaxies contained within the thin black lines, or $\pm 1\sigma$), the mean merger mass ratio is essentially independent of stellar mass (Fig. \ref{fig:merging_history}, right panel). In other words, the `isochromes' in the two-dimensional histogram are parallel to the median trend as a function of stellar mass. This is consistent with \cite{Rodriguez-Gomez2015}, where it was found that the mass ratio dependence of the galaxy merger rate is approximately universal.

Among the three different merger statistics described above, perhaps the one that has been the topic of most discussion in the literature, at least regarding its effects on galaxy morphology, is the gas content of the merger, quantified here as the mean merger gas fraction. It is generally accepted that dry mergers are more effective at creating spheroids than gas-rich mergers \citep[e.g.][]{Khochfar2003, Naab2006a, Hopkins2010}.

Fig. \ref{fig:kappa_vs_merger_gas_fraction} shows the correlation between galaxy morphology and mean merger gas fraction, using the same plotting style as Fig. \ref{fig:kappa_vs_f_acc}. This figure shows that gas-poor mergers are indeed more effective spheroid formation mechanisms than gas-rich mergers, although the trend is significant only for galaxies with $M_{\ast} \gtrsim 10^{11} \Msun$. This corresponds to the mass range where the fraction of accreted stellar mass has an impact on galaxy morphology (compare to Fig. \ref{fig:kappa_vs_f_acc}), which is perhaps unsurprising given the strong, positive correlation between $f_{\rm acc}$ and the mean merger gas fraction displayed in Fig. \ref{fig:merging_history}. These results are consistent with those of \cite{Lagos2017a}, who found that dry mergers have a more dramatic impact than gas-rich mergers on the angular momentum content of galaxies. The effects of gas fraction on individual mergers of Illustris galaxies is explored in detail in Penoyre et al. (in prep.), while \cite{Sparre2016} investigate the evolution of galaxy morphology during and after gas-rich mergers using high-resolution `zoom-in' resimulations of Illustris mergers.

We highlight that Fig. \ref{fig:kappa_vs_merger_gas_fraction} is showing the gas fraction of the secondary progenitors, i.e. the gas brought in by mergers. Therefore, one might wonder about the effect on morphology due to gas in the primary galaxy itself. To investigate this, we compared $\kappa_{\rm rot}$ to the gas fraction along a galaxy's main branch\footnote{Defined as $\sum{M_{\rm gas}} / \sum{\left(M_{\rm gas} + M_{\ast}\right)}$, where the sum is carried out over the galaxy's main branch in the merger tree.} and found the same qualitative behaviour as in Fig. \ref{fig:kappa_vs_merger_gas_fraction}: $\kappa_{\rm rot}$ is correlated with gas fraction in massive galaxies, but the trend weakens at $M_{\ast} \lesssim 10^{11} \Msun$. This result, which we do not show for the sake of brevity, can be understood by noting that most galaxies with stellar masses below $M_{\ast} \approx 10^{11} \Msun$ have been gas-rich throughout their histories, and therefore they all have the same `ability' to form new stars and regrow their stellar discs. This allows other factors to play a more important role in determining the morphology of smaller galaxies, as we discuss in the next section.

\subsection{Halo spin}\label{subsec:halo_spin}

Now we test whether the spin of a DM halo has an important effect on the morphology of the galaxy formed at its centre. For instance, the analytic arguments presented in \cite{Fall1983} and \cite{Mo1998} suggest that galactic discs that form in haloes with large spin should have large sizes and low surface densities, a prediction that has been shown to work reasonably well in cosmological hydrodynamic simulations \citep{Sales2009, Grand2016}. However, a systematic comparison between the spin of the DM halo and the $\kappa_{\rm rot}$ parameter over a wide range of stellar masses has not been done before.

We parametrize halo spin using the definition from \cite{Bullock2001},
\begin{equation}
\lambda^{\prime} \equiv \frac{J_{\rm 200}}{\sqrt{2} M_{\rm 200} V_{\rm 200} R_{\rm 200}},
\label{eq:spin_parameter}
\end{equation}
where $R_{\rm 200} \equiv R_{\rm 200, crit}$ is the radius enclosing an average density equal to 200 times the critical density of the Universe, $M_{\rm 200} \equiv M_{\rm 200, crit}$ is the total mass within $R_{\rm 200}$, $J_{\rm 200}$ is the magnitude of the total angular momentum within $R_{\rm 200}$, and $V_{\rm 200} = \sqrt{G M_{\rm 200} / R_{\rm 200}}$. For our analysis we use the angular momentum vectors from the Illustris halo catalogue extension presented in \cite{Zjupa2017}. In this extension, $J_{\rm 200}$ is calculated including all resolution elements contained within $R_{\rm 200}$, rather than only those which are also members of the FoF group in question.

The halo spin parameter defined by equation (\ref{eq:spin_parameter}) is approximately mass-independent and its statistical distribution is well fitted by a log-normal distribution \citep[e.g.][]{Barnes1987, VandenBosch1998, Bullock2001},
\begin{equation}
{\rm P}(\lambda^{\prime}) = \frac{1}{\lambda^{\prime}\sqrt{2\pi}\sigma} \exp\left(-\frac{\ln^2(\lambda^{\prime}/\lambda^{\prime}_0)}{2\sigma^2}\right),
\label{eq:log-normal}
\end{equation}
where $\lambda^{\prime}_0$ and $\sigma$ are the best-fitting values.\footnote{A slightly more accurate fitting formula was later proposed by \cite{Bett2007}.} A detailed study of halo spin in the Illustris and Illustris-Dark simulations has already been carried out in \cite{Zjupa2017}.

In order to study the effect of halo spin on the morphology of the galaxy hosted at its centre, while at the same time removing the influence of the baryons themselves on halo spin, we match each subhalo from Illustris to its counterpart in Illustris-Dark as described in Section \ref{subsec:matched_catalogues}. We denote these DMO-matched halo spin parameters by $\lambda^{\prime}_{\rm DMO}$, and we will use them exclusively throughout the rest of this paper. We note, however, that none of our results would change appreciably if we had used the halo spin parameters from the hydrodynamic simulation.

\begin{figure}
  \centerline{\hbox{
	\includegraphics[width=8cm]{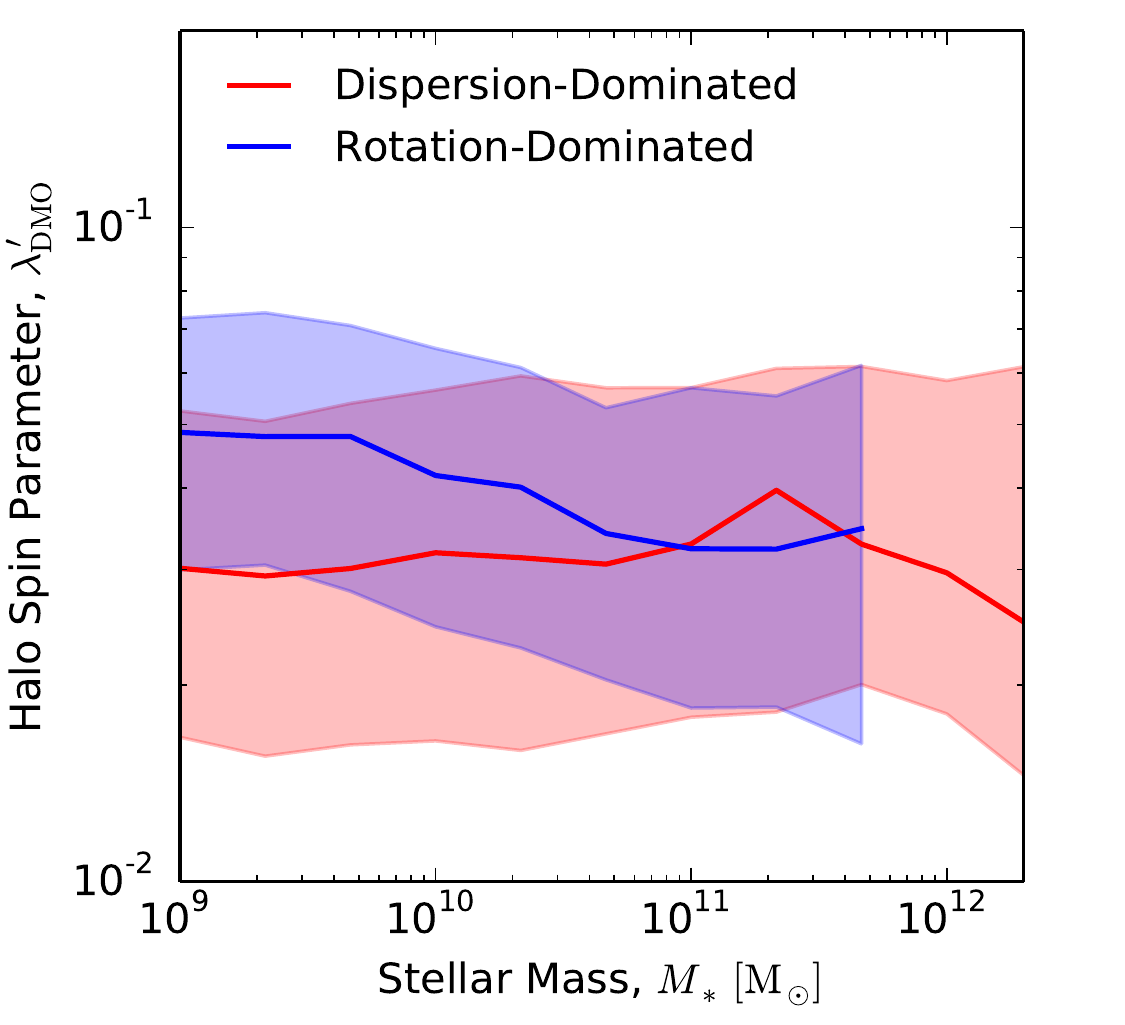}
  }}
	\caption{The median halo spin parameter (matched to an analogous DMO simulation) as a function of stellar mass, shown for central galaxies at $z=0$. As before, the galaxies have been separated into two morphological types according to their $\kappa_{\rm rot}$ values, with red and blue lines representing dispersion- and rotation-dominated systems, respectively. The shaded regions indicate the 16th to 84th percentile ranges.}
	\label{fig:spin_parameter_vs_mstar}
\end{figure}

\begin{figure*}
  \centering
	\includegraphics[width=17.5cm]{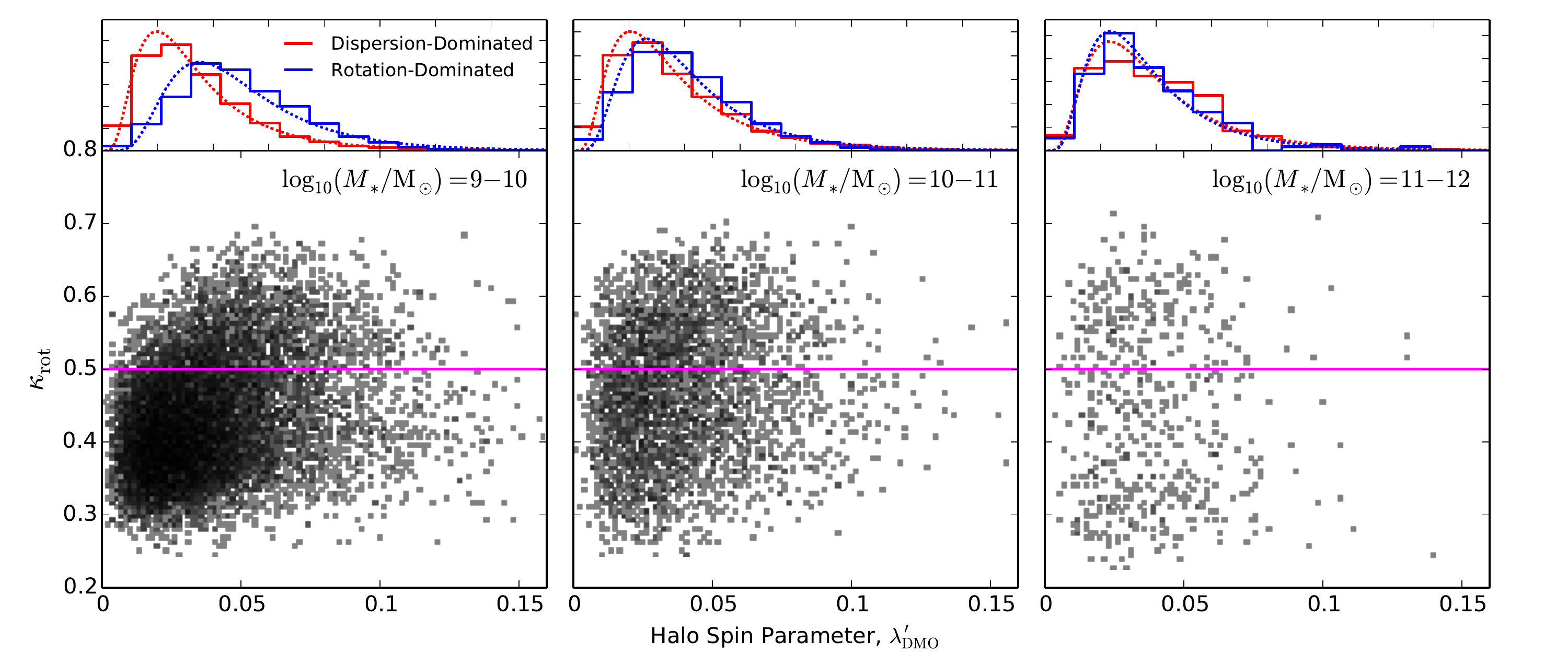}
	\caption{Correlation between galaxy morphology and halo spin parameter (matched to an analogous DMO simulation) at $z=0$. The scatter plots in the lower panels show the overall galaxy distribution, while the histograms in the upper panels show the corresponding marginal distributions of dispersion- and rotation-dominated systems. The dotted lines in the upper panels show the best-fitting log-normal distributions, with best-fitting parameters given in Table \ref{tab:log_normal_fits}.}
	\label{fig:kappa_vs_spin_parameter}
\end{figure*}

Fig. \ref{fig:spin_parameter_vs_mstar} shows the median spin parameter of DM haloes as a function of the stellar mass of the galaxies hosted at their centres, distinguishing between dispersion-dominated ($\kappa_{\rm rot} < 0.5$; shown in red) and rotation-dominated ($\kappa_{\rm rot} \geq 0.5$; shown in blue) systems. The shaded regions indicate the 16th to 84th percentile ranges of the individual distributions. This figure shows that the effect of halo spin on galaxy morphology is mass-dependent, being most effective in smaller galaxies ($M_{\ast} \lesssim 10^{10} \, \Msun$) and becoming progressively weaker in more massive systems.

In Fig. \ref{fig:kappa_vs_spin_parameter} we further explore the correlation between morphology and halo spin. As before, the two-dimensional histograms (coloured using the same range of greys as in Fig. \ref{fig:morphology_vs_mass}) are effectively scatter plots showing the overall galaxy distribution in different stellar mass bins, while the histograms in the upper panels show the marginal distributions for dispersion- and rotation-dominated systems. The data in the upper panels have been fitted with a log-normal distribution, with best-fit parameters specified in Table \ref{tab:log_normal_fits}. 

The left-hand panels of Fig. \ref{fig:kappa_vs_spin_parameter} once again show that there is a correlation between galaxy morphology and halo spin: galaxies with higher (lower) $\kappa_{\rm rot}$ are more likely to be found at the centres of haloes with higher (lower) spin parameter, although there is significant overlap in the $\lambda^{\prime}$ distributions of dispersion-dominated and rotation-dominated systems. This trend becomes weaker at $M_{\ast} \gtrsim 10^{10} \Msun$ (middle and right panels), in agreement with several previous results in this mass range \citep[e.g.][]{Scannapieco2009, Sales2012}.

Recently, however, \cite{Teklu2015} studied the halo spin parameter of disc- and spheroid-dominated galaxies with $M_{\rm halo} > 5 \times 10^{11} \Msun$ (approximately $M_{\ast} \gtrsim 10^{10} \Msun$), finding a mild preference for disc-dominated galaxies to form in haloes with larger spin (see their figure 15). This is unlikely to be in contradiction with our work or with previous ones. In fact, their trend is similar to the one found in our sample, but only when we consider galaxies with $M_{\ast} \lesssim 10^{10} \Msun$. Because low-mass galaxies dominate in number compared to more massive systems, the signal prevails when the whole galaxy sample is considered at stellar masses $M_{\ast} \gtrsim 10^{10} \Msun$, which is the case in \cite{Teklu2015}. Nevertheless, we highlight that, at least in our sample, the trend between morphology and halo spin does exist in dwarf galaxies, disappearing later for more massive systems, likely as a combination of more complex assembly histories and the increasingly important role of mergers.

\subsection{The complementary roles of mergers and halo spin}\label{subsec:mergers_and_halo_spin}

\begin{table}
	\centering
	\begin{tabular}{c | c | c}
  	\hline
  	$M_{\ast}$ range & Dispersion-dominated & Rotation-dominated \\
  	\hline
    $10^{9}$--$10^{10} \, \Msun$ & $\lambda_{0}^{\prime} = 0.029$, $\sigma = 0.62$ & $\lambda_{0}^{\prime} = 0.045$, $\sigma = 0.50$ \\
		\addlinespace[1.5ex]
    $10^{10}$--$10^{11} \, \Msun$ & $\lambda_{0}^{\prime} = 0.031$, $\sigma = 0.63$ & $\lambda_{0}^{\prime} = 0.036$, $\sigma = 0.55$ \\
		\addlinespace[1.5ex]
    $10^{11}$--$10^{12} \, \Msun$ & $\lambda_{0}^{\prime} = 0.033$, $\sigma = 0.60$ & $\lambda_{0}^{\prime} = 0.032$, $\sigma = 0.56$ \\
  	\hline
	\end{tabular}
  \caption{Best-fitting parameters of the log-normal fits shown in Fig. \ref{fig:kappa_vs_spin_parameter}.}
  \label{tab:log_normal_fits}
\end{table}

\begin{figure*}
  \centering
	\includegraphics[width=17.5cm]{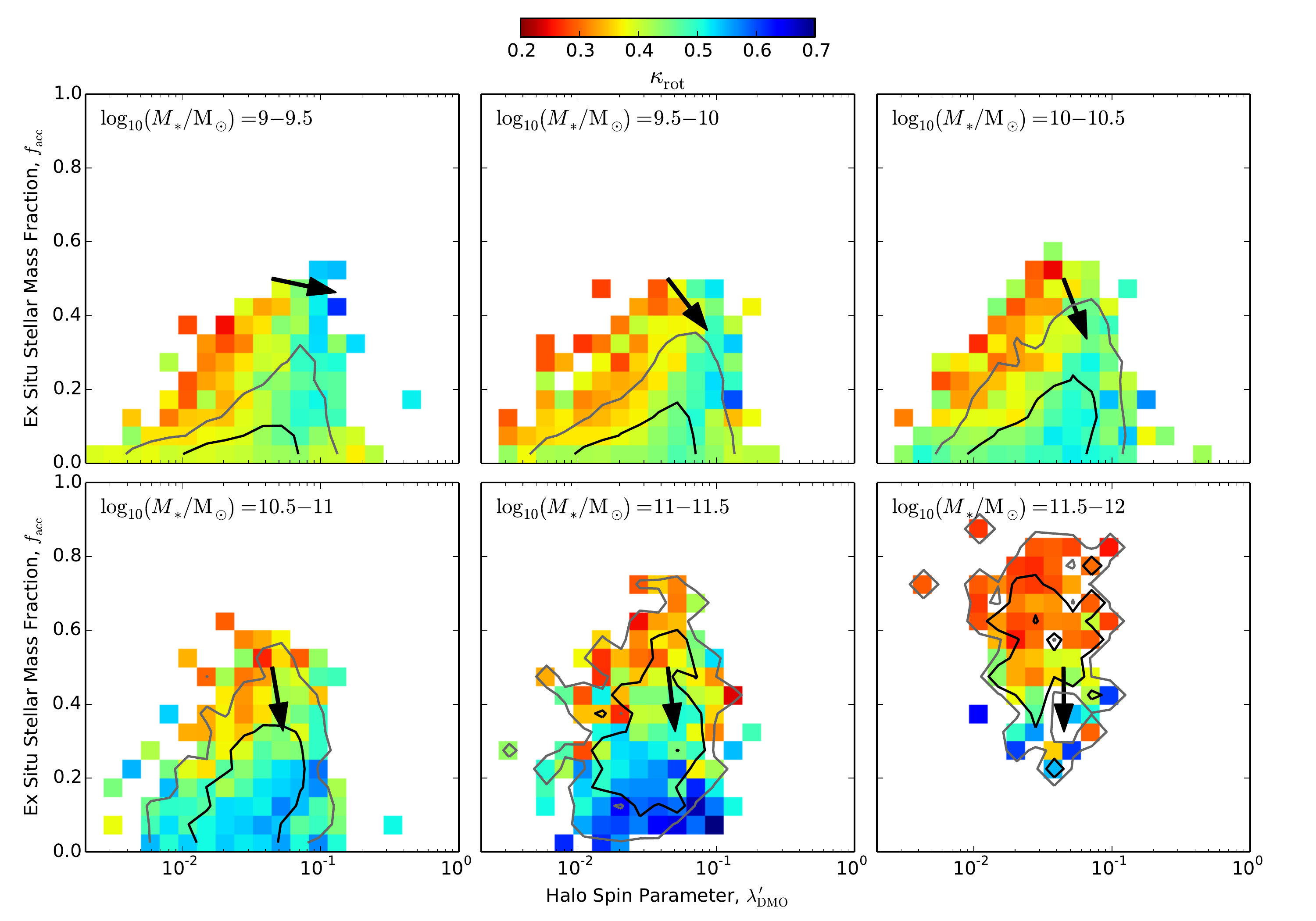}
	\caption{Dependence of galaxy morphology on the halo spin parameter ($x$-axis) and the \textit{ex situ} stellar mass fraction ($y$-axis), shown for central galaxies at $z=0$ in different stellar mass bins (different panels). The two-dimensional histograms are coloured according to the median value of $\kappa_{\rm rot}$ of the galaxies that fall into each two-dimensional bin. The black arrows indicate the direction of increasing $\kappa_{\rm rot}$, while the contours contain 68 and 95 per cent of the galaxy population in each stellar mass bin. At low stellar masses (upper-left panel), the $\kappa_{\rm rot}$ gradient becomes almost horizontal, while at high stellar masses (bottom-right panel) the $\kappa_{\rm rot}$ gradient becomes approximately vertical. For intermediate masses (second to fifth panels), galaxy morphology is correlated with a combination of the \textit{ex situ} stellar mass fraction and the halo spin parameter, such that the $\kappa_{\rm rot}$ gradient becomes approximately diagonal. The variation in the direction of this $\kappa_{\rm rot}$ gradient once again shows that galaxy morphology is mostly determined by halo spin at low masses, while mergers have a dominant effect on the morphology of more massive galaxies.}
	\label{fig:morphology_diagram_manymasses}
\end{figure*}

Here we expand on our results from Sections \ref{subsec:merging_history} and \ref{subsec:halo_spin}, but now focusing on the idea that merging history and halo spin play complementary roles in shaping galaxy morphology. We know from Fig. \ref{fig:kappa_vs_f_acc} that mergers are more effective at shaping the morphology of massive galaxies ($M_{\ast} = 10^{11}$--$10^{12} \Msun$), while Fig. \ref{fig:kappa_vs_spin_parameter} suggests that halo spin plays a more important role at lower masses ($M_{\ast} = 10^{9}$--$10^{10} \Msun$). However, the effect from any of these two factors \textit{alone} is relatively unimportant for medium-sized galaxies ($M_{\ast} = 10^{10}$--$10^{11} \Msun$). Here we explore the idea that a combination of \textit{both} merging history and halo spin could be useful in understanding what determines the morphology of galaxies with masses similar to that of the Milky Way.

Fig. \ref{fig:morphology_diagram_manymasses} shows how the morphologies of central galaxies at $z=0$ depend on halo spin ($x$-axis), \textit{ex situ} stellar mass fraction ($y$-axis), and stellar mass (different panels). Each two-dimensional histogram is coloured according to the median $\kappa_{\rm rot}$ value of the galaxies that fall into each two-dimensional bin, while the black and dark grey contours contain 68 and 95 per cent of the galaxies in each panel. The black arrows indicate the direction of increasing $\kappa_{\rm rot}$, obtained from a linear fit to the data in each panel, weighted by the number of galaxies.\footnote{We excluded galaxies with $f_{\rm acc} < 0.02$, since they are not well described by a plane in the upper left panel of Fig. \ref{fig:morphology_diagram_manymasses}. This has negligible effects in the other five panels, i.e. for galaxies with $M_{\ast} > 10^{9.5} \Msun$.} This figure shows that, in general, $\kappa_{\rm rot}$ depends on a combination of halo spin and \textit{ex situ} fraction, as manifested by the `diagonal' $\kappa_{\rm rot}$ gradient seen in most panels: galaxies with violent (quiet) merging histories which are also found at the centres of slowly (fast) rotating haloes are more likely to be dispersion-dominated (rotation-dominated).

However, we had demonstrated in Sections \ref{subsec:merging_history} and \ref{subsec:halo_spin} that the effect of merging history and halo spin on galaxy morphology depends on the stellar mass of the galaxies considered. This mass-dependent behaviour is manifested in Fig. \ref{fig:morphology_diagram_manymasses} by the fact that the direction of the $\kappa_{\rm rot}$ gradient changes with stellar mass. At low masses (top-left panels), the $\kappa_{\rm rot}$ gradient is almost horizontal, which implies that the morphology of galaxies with `similar' merging histories (i.e. with the same $f_{\rm acc}$ value) is mostly determined by the spin parameter of their parent haloes. On the other hand, the $\kappa_{\rm rot}$ gradient becomes approximately vertical at $M_{\ast} \gtrsim 10^{11} \Msun$ (bottom-right panels), which means that mergers have a major impact on the morphology of massive galaxies, erasing any `pre-existing dependence' on halo spin.

In order to interpret the changing direction of such gradients in Fig. \ref{fig:morphology_diagram_manymasses} correctly, one must consider the following factors. Firstly, the stellar mass of a galaxy is strongly correlated with its gas fraction, as well as with the gas fraction of other galaxies that have merged onto it (Fig. \ref{fig:merging_history}, left panel). In general, more gas is involved in galaxy interactions at low stellar masses. Secondly, $f_{\rm acc}$ does not measure the merging history \textit{per se}, but rather the impact of dry mergers relative to \textit{in situ} star formation. Therefore, a low-mass galaxy with high $f_{\rm acc}$ (an indication of a recent, gas-rich major merger, as implied by Fig. \ref{fig:merging_history}) will nevertheless possess a considerable amount of gas, a higher specific star formation rate, and therefore the `ability' to regrow its stellar disc (a quite direct demonstration of this effect was recently presented in \citealt{Sparre2016}). The $\kappa_{\rm rot}$ values of these low-mass discs are ultimately determined by the halo spin. Finally, there might be other factors driving the mass-dependent trends in Fig. \ref{fig:morphology_diagram_manymasses}, but whose effects are more challenging to quantify. For example, it has been argued that a large fraction of the gas accreted onto small galaxies at $z<1$ comes from a `galactic fountain' of recycled gas \citep[e.g.][and references therein]{Nelson2015a}. This effect could introduce or enhance correlations between the angular momentum of the halo and that of the galaxy formed at its centre \citep[e.g.][]{Genel2015}.

We finalize this section by making a more subtle observation about Fig. \ref{fig:morphology_diagram_manymasses}, which is that the \textit{ex situ} stellar mass fraction $f_{\rm acc}$ and the halo spin parameter $\lambda^{\prime}_{\rm DMO}$, usually assumed to be independent from each other, actually appear to be somewhat correlated, as evidenced by the mildly diagonal orientation of the contour lines. This suggests that galaxy mergers are associated with higher halo spin, as previously speculated by a number of investigations (e.g. \citealt{Vitvitska2002, Peirani2004, Hetznecker2006, Bett2012, Welker2014}; but see \citealt{DOnghia2007}, for an interesting caveat), although it is unclear whether this correlation implies causation (i.e. mergers directly increasing the spin of the DM halo), or whether haloes with higher spin simply happen to form in environments which are also favourable to galaxy mergers, such as along galaxy filaments.\footnote{The high velocity dispersion found in cluster environments, along with the associated lower gas fractions, tends to lower the likelihood of two nearby galaxies in such dense environments actually merging.} A full discussion of this curious effect is beyond the scope of this paper.

\section{Discussion and conclusions}\label{sec:discussion_and_conclusions}

We have investigated the connection between galaxy morphology, halo spin, and merging history in the Illustris cosmological simulation, considering approximately 18,000 central galaxies at $z=0$ over a wide range of stellar masses ($M_{\ast} = 10^{9}$--$10^{12} \Msun$). We showed that disc-like and spheroidal galaxies arise naturally in the Illustris simulation over the mass range considered (Figs. \ref{fig:disks} and \ref{fig:spheroids}), and then proceeded to investigate the effects of mergers and halo spin on galaxy morphology as a function of stellar mass.

Throughout this work, we quantified galaxy morphology using the $\kappa_{\rm rot}$ parameter \citep{Sales2012}, defined in equation (\ref{eq:kappa}). This quantity is closely related to other kinematic morphological parameters, such as the kinematic disc-to-total fraction (sometimes denoted by D/T) that is often used in other simulation studies (see Section \ref{subsec:morphology_definitions}). Based on the arguments presented in Appendix \ref{app:kappa_expanded}, we employed the $\kappa_{\rm rot}$ parameter to classify galaxies as rotation-dominated ($\kappa_{\rm rot} \geq 0.5$) or dispersion-dominated ($\kappa_{\rm rot} < 0.5$), as illustrated in Fig. \ref{fig:morphology_vs_mass}.

In order to test the idea that mergers play an important role in determining galaxy morphology \citep[e.g.][]{Toomre1977, White1978, Barnes1996, Naab2006, Naab2014}, we used the \textit{ex situ} stellar mass fraction $f_{\rm acc}$ to measure the overall impact of galaxy mergers (relative to \textit{in situ} star formation). We found that mergers are indeed an important transformational mechanism in massive galaxies ($M_{\ast} \gtrsim 10^{11} \Msun$), but their importance diminishes at lower masses (Figs. \ref{fig:f_acc_vs_mstar} and \ref{fig:kappa_vs_f_acc}). To the best of our knowledge, this is the first time that a statistical demonstration of the `merger hypothesis' has been carried out with a cosmological hydrodynamic simulation over such a wide range of stellar masses (however, see \citealt{DeLucia2011} for a study using SAMs).

Beyond the \textit{ex situ} stellar mass fraction, $f_{\rm acc}$, we also considered three additional merger statistics: the mean merger gas fraction, the mean merger lookback time, and the mean merger mass ratio. By comparing these quantities to $f_{\rm acc}$ we found that -- at any fixed stellar mass -- a higher $f_{\rm acc}$ is associated with a larger number of \textit{massive}, \textit{recent}, and \textit{dry} mergers (Fig. \ref{fig:merging_history}), which happen to be the kind of mergers that are believed to contribute most to the formation of spheroidal galaxies. In fact, in Fig. \ref{fig:kappa_vs_merger_gas_fraction} we showed explicitly that gas-poor mergers contribute more to spheroid formation than gas-rich mergers, although the trend is strong only for massive galaxies ($M_{\ast} \gtrsim 10^{11} \Msun$).

Having investigated the role of merging history in shaping galaxy morphology, we proceeded to test the hypothesis that, through conservation of specific angular momentum, the spin of a halo also plays a major role in determining the morphology of the galaxy formed at its centre. According to this picture, DM haloes with high angular momentum content favour the formation of rotationally supported discs. In order to rule out baryonic effects on halo spin, we matched each Illustris halo to its counterpart in Illustris-Dark and used the spin parameter of the latter for all our analyses. We showed in Figs. \ref{fig:spin_parameter_vs_mstar} and \ref{fig:kappa_vs_spin_parameter} that $\kappa_{\rm rot}$ is somewhat determined by halo spin at $M_{\ast} \lesssim 10^{10} \Msun$, but the correlation between these two quantities becomes weaker at higher masses and eventually disappears, presumably due to the increasing impact of galaxy mergers.

Finally, we investigated the joint effect of halo spin and merging history on galaxy morphology. We found that, in general, galaxies with high (low) $f_{\rm acc}$ which are also located at the centres of slowly (fast) rotating haloes are more likely to be dispersion-dominated (rotation-dominated) systems, as manifested by the `diagonal' $\kappa_{\rm rot}$ gradients in Fig. \ref{fig:morphology_diagram_manymasses}. However, the relative importance between halo spin and $f_{\rm acc}$ is mass-dependent, as evidenced by the varying \textit{direction} of the $\kappa_{\rm rot}$ gradients: halo spin is the dominant driver of galaxy morphology at low masses, even for galaxies which have undergone major mergers, while dry mergers are more important in determining the morphology of more massive systems. The morphology of Milky Way-sized galaxies, at the transition between these two regimes ($M_{\ast} \approx 10^{10}$--$10^{11} \, \Msun$), depends on a combination of $f_{\rm acc}$ and halo spin, as described above, but shows a weak dependence on any of these two factors alone. We argued that the physical origin of this mass-dependent response of galaxy morphology to mergers is largely due to the nature of the mergers undergone by galaxies in different mass ranges.

Indeed, as discussed in Section \ref{subsec:merging_history}, two galaxies of different masses with the same $f_{\rm acc}$ value need not have `equivalent' merging histories. For example, Fig. \ref{fig:merging_history} shows that low-mass galaxies are more likely to have gas-rich mergers, while the merging history of a more massive system usually consists of numerous dry mergers. This has crucial consequences for galaxy morphology. In fact, it has been proposed that dry mergers are important in the formation of spheroidal systems \citep[e.g.][]{Khochfar2003, Naab2006a}, while gas-rich major mergers have been shown to produce disc galaxies \citep[e.g.][]{Springel2005, Robertson2006, Governato2007}. Both of these ideas seem to be manifested statistically in the different panels of Fig. \ref{fig:morphology_diagram_manymasses}, where the `likelihood' of galaxy mergers producing a spheroid increases with stellar mass, along with the `dryness' of the galaxy mergers in each mass range (Fig. \ref{fig:merging_history}). Furthermore, we know that low-mass galaxies undergo major mergers somewhat less frequently than massive galaxies, by about an order of magnitude \citep{Rodriguez-Gomez2015}. This reduced merger frequency, along with the higher gas fractions (and specific star formation rates) present at low stellar masses, further adds to the apparent `resilience' of small galaxies to the destructive effects of major mergers.

In other words, we interpret the shifting importance of mergers in determining galaxy morphology as being largely due to the varying nature of galaxy mergers in different mass ranges: mergers between low-mass galaxies tend to be gas-rich and less frequent, while the merging history of a massive galaxy is dominated by repeated gas-poor mergers. In addition, low-mass galaxies have larger gas reservoirs which allow them to regrow their stellar discs. Thus, our work supports a scenario in which galaxy mergers manifest themselves as different kinds of transformational mechanisms, potentially giving rise to both discs and spheroids, depending on the mass range considered.

\section*{Acknowledgements}

We thank Nicola Amorisco, Martin Sparre, Julio Navarro, Michael Fall and Adrian Jenkins for useful comments and discussions, as well as Rachel Somerville for carefully reading the manuscript. SG and PT acknowledge support provided by NASA through Hubble Fellowship grants HST-HF2-51341.001-A and HST-HF2-51384.001-A awarded by the STScI, which is operated by the Association of Universities for Research in Astronomy, Inc., for NASA, under contract NAS5-26555. JZ acknowledges support from the Deutsche Forschungsgemeinschaft through Transregio 33, `The Dark Universe,' and from IMPRS for Astronomy and Cosmic Physics at the University of Heidelberg. JZ and VS would like to thank the Klaus Tschira Foundation. VS also acknowledges support through the European Research Council through ERC-StG grant EXAGAL-308037. GS appreciates support from a Giacconi Fellowship at the STScI, which is operated by the Association of Universities for Research in Astronomy, Inc., under NASA contract NAS 5-26555. LH acknowledges support from NSF grant AST-1312095 and NASA grant NNX12AC67G. Simulations were run on the Harvard Odyssey and CfA/ITC clusters, the Ranger and Stampede supercomputers at the Texas Advanced Computing Center as part of XSEDE, the Kraken supercomputer at Oak Ridge National Laboratory as part of XSEDE, the CURIE supercomputer at CEA/France as part of PRACE project RA0844, and the SuperMUC computer at the Leibniz Computing Centre, Germany, as part of project pr85je. The Flatiron Institute is supported by the Simons Foundation.

\bibliographystyle{mn2eFixed}

\bibliography{paper}

\appendix

\section{The kappa morphological parameter}\label{app:kappa_expanded}

This appendix contains an expanded discussion about the $\krot$ morphological parameter, which was introduced in equation (\ref{eq:kappa}). In particular, we calculate $\krot$ values for a few idealized, analytically tractable stellar systems, and then show that $\krot$ can be used as a proxy for rotational support in our simulated galaxies.

Since the $z$-component of a particle's specific angular momentum is $j_z = R v_{\phi}$, where $R$ is the distance to the $z$-axis and $v_{\phi}$ is the azimuthal component of the velocity, $\krot$ can be conveniently expressed as

  \begin{equation}
    \kappa_{\rm rot} = \frac{\sum_{i} m_{i} v_{\phi,i}^{2}}{\sum_{i} m_{i} v_{i}^{2}},
    \label{eq:kappa2}
  \end{equation}
where $m_i$ and $v_i$ represent mass and total velocity, respectively, and the sum is carried out over all stellar particles. Expression (\ref{eq:kappa2}) shows that, by construction, $\krot$ is similar to other kinematic morphological parameters such as $V/\sigma$ \citep[e.g.][]{Dubois2016}, except that $\krot$ does not distinguish the direction of rotation about the $z$-axis (i.e. co-rotating or counter-rotating), as discussed later in this appendix.

Furthermore, if we model a stellar system as being composed of infinitely many particles and having a smooth density distribution $\rho({\bf r})$, where ${\bf r}$ denotes position with respect to the galactic centre, then $\krot$ can also be expressed as

\begin{equation}
\krot = \frac{\int \rho({\bf r}) \left<v_{\phi}^2({\bf r})\right> {\rm d}{\bf r}}{\int \rho({\bf r}) \left<v^2({\bf r})\right> {\rm d}{\bf r}},
\label{eq:kappa_int}
\end{equation}
where $\left<v_{\phi}^2({\bf r})\right>$ and $\left<v^2({\bf r})\right>$ are averaged over particles found within a small volume around position ${\bf r}$. Note that, for a spherically symmetric system, equation (\ref{eq:kappa_int}) becomes

\begin{equation}
\krot = \frac{\int_0^{\infty} \rho(r) \left<v_{\phi}^2(r)\right> r^2 {\rm d}r}{\int_0^{\infty} \rho(r) \left<v^2(r)\right> r^2 {\rm d}r},
\label{eq:kappa_int_spherical}
\end{equation}
where $r \equiv \left|{\bf r}\right|$ is the distance to the galactic centre.

For illustrative purposes, below we calculate $\krot$ for some simple dynamical systems, some of which may not exactly correspond to real or simulated galaxies.

\begin{itemize}
  \item \textit{Highly eccentric orbits} ($\krot = 0$): A system with highly eccentric orbits has $\left<v_{r}^2\right> \gg \left<v_{\phi}^2\right>$ and $\left<v_{r}^2\right> \gg \left<v_{\theta}^2\right>$, where $v_{r}$ and $v_{\theta}$ are the the radial and polar velocity components, respectively (i.e. $v_{\theta} = r\dot{\theta}$, where $\theta$ is the polar angle). In this system, stars in radial orbits dominate the kinetic energy but do not contribute to the angular momentum, so in the limiting case of purely radial orbits we have $\krot = 0$. In theory, such a system could arise from the collapse of an isolated, non-rotating cloud of gas. However, these singular systems do not exist in Illustris, as demonstrated in Section \ref{subsec:morphology_distribution}. \\

  \item \textit{Dispersion-supported, isotropic stellar system} ($\krot = 1/3$): Consider a spherically symmetric system where all three velocity components have the same statistical distribution at any radius $r$. Then, for a randomly chosen $z$-axis, we have $\left<v_{\phi}^2(r)\right> = \frac{1}{3}\left<v^2(r)\right>$. Substituting into equation (\ref{eq:kappa_int_spherical}) yields $\krot = 1/3$. As shown in Section \ref{subsec:morphology_distribution}, some Illustris galaxies have $\krot$ values in this range, while some of them have even smaller values ($\krot \lesssim 0.3$), presumably due to a combination of high velocity dispersion and somewhat eccentric orbits. \\

  \item \textit{Cold rotating disc} ($\krot = 1$): Finally, a stellar system where all particles move in circular orbits about the $z$-axis has $\left<v_{\phi}^2({\bf r})\right> = \left<v^2({\bf r})\right>$ everywhere. Therefore, from equation (\ref{eq:kappa_int}), this system has $\krot = 1$. Note that this is only valid for a thin, dynamically cold disc. In general, stars belonging to a thick, dynamically hot disc oscillate along the $z$-direction, which would lead to $\krot < 1$. Galactic discs in Illustris are somewhat thick, reaching maximum $\krot$ values around 0.7. Also note that a cold, disc-like system where roughly half of the stars rotate in the direction of the total angular momentum, while the other half counter-rotate, would also have $\krot = 1$ (and yet almost no net rotation). Although mathematically possible, this configuration does not seem common in cosmological simulations, as shown below.

\end{itemize}

\begin{figure}
  \centering
	\includegraphics[width=8cm]{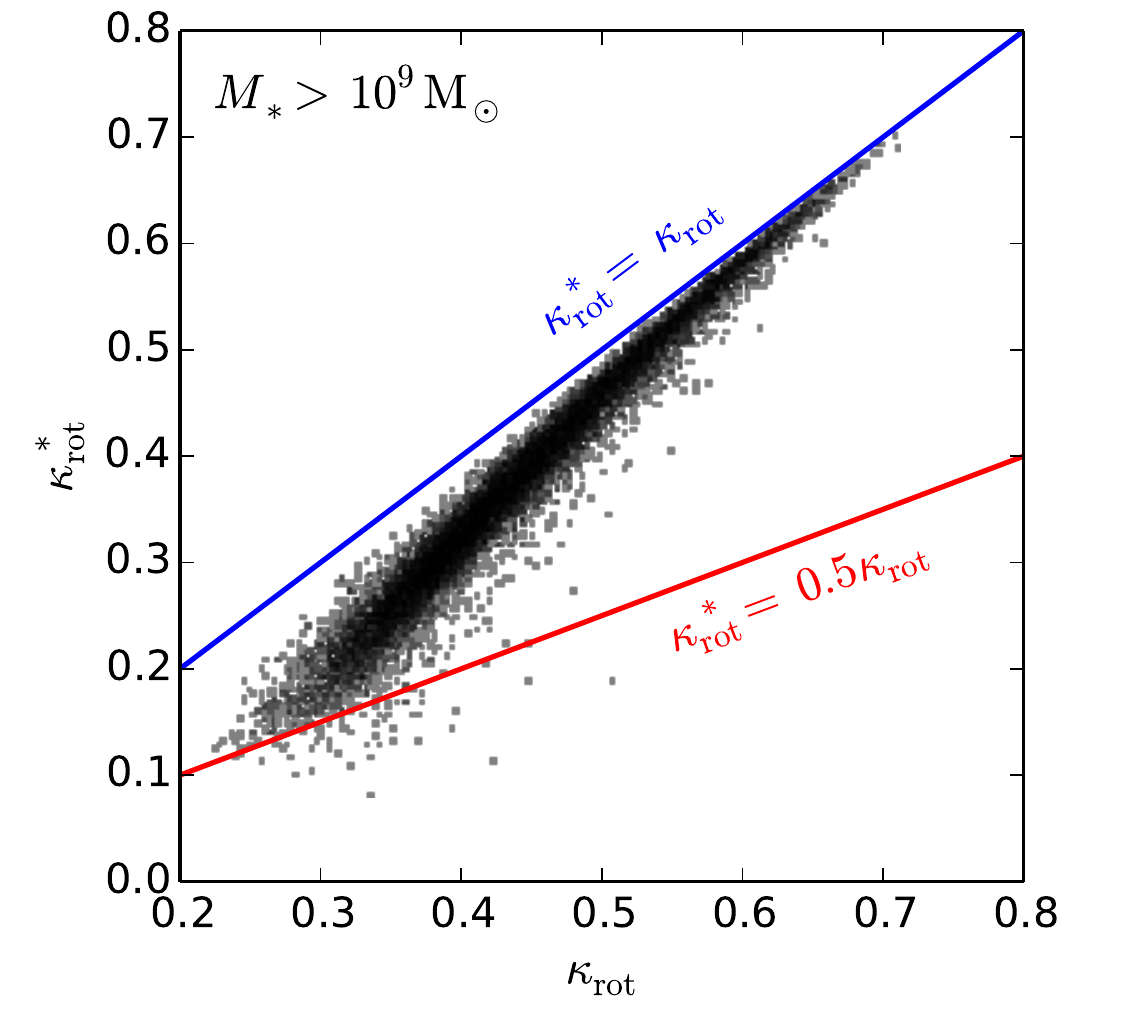}
	\caption{Correlation between $\kappa_{\rm rot}^{\ast}$ and $\krot$ for central galaxies at $z=0$ with $M_{\ast} > 10^{9} \, \Msun$. For reference, the blue and red lines correspond to $\kappa_{\rm rot}^{\ast} = \kappa_{\rm rot}$ and $\kappa_{\rm rot}^{\ast} = 0.5 \kappa_{\rm rot}$, respectively. The Pearson correlation coefficient of the data is $r = 0.985$.}
	\label{fig:kappa_vs_kappa_star}
\end{figure}

Because of the sign ambiguity illustrated in the last point, $\krot$ does not formally measure ordered rotation along the $z$-axis. In practice, however, for cosmologically simulated galaxies $\krot$ turns out to be an excellent proxy for ordered rotation.\footnote{See also figure 1 in \cite{Sales2012}.} To demonstrate this, we consider an alternative quantity $\kappa_{\rm rot}^{\ast}$, which is also defined by equation (\ref{eq:kappa}) except that the sum is only carried out over stars that circulate in the same direction as the total stellar angular momentum (namely, stars that have $j_z > 0$). The denominator $K$ is unchanged (i.e. all stars contribute to the total kinetic energy).

Fig. \ref{fig:kappa_vs_kappa_star} shows the relationship between $\krot$ and $\kappa_{\rm rot}^{\ast}$ for the Illustris galaxy sample, with a greyscale showing the overall galaxy distribution. Encouragingly, what we consider disc-dominated systems in this paper turn out to be truly rotation-supported objects, with $\kappa_{\rm rot}^{\ast} \approx \kappa_{\rm rot}$ (since none or very few stars counter-rotate), while dispersion-supported systems have $\kappa_{\rm rot}^{\ast} \approx 0.5 \kappa_{\rm rot}$ (since roughly half of the stars counter-rotate).

More importantly, Fig. \ref{fig:kappa_vs_kappa_star} shows that $\krot$ and $\kappa_{\rm rot}^{\ast}$ are very strongly correlated in our simulated galaxies, with a Pearson correlation coefficient of $r = 0.985$. In previous work, \cite{Sales2012} showed that $\krot$ is strongly correlated with the fraction of stars in corotating circular orbits (their figure 1). All of this justifies the use of $\krot$ as a proxy for rotational support in a galaxy, as we do throughout this paper.

\section{Galaxy morphology and \\ stellar angular momentum}\label{app:angular_momentum}

Since $\kappa_{\rm rot}$ (or, similarly, the disc-to-total fraction D/T) and the specific angular momentum $j_{\ast}$ are both associated with the rotation of the galactic stellar component, and yet in the main body of this paper we focused only on the former, it is beneficial to investigate what are exactly the differences between these two quantities, as we briefly do in this appendix.

\begin{figure*}
  \centering
	\includegraphics[width=17.5cm]{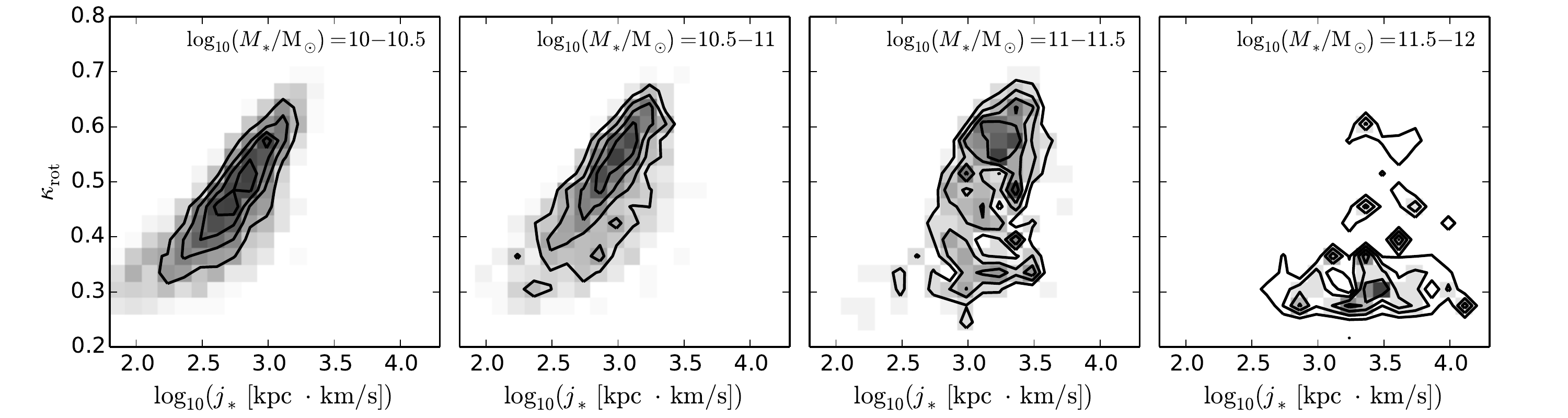}
	\caption{Correlation between $\kappa_{\rm rot}$ and specific angular momentum $j_{\ast}$ for galaxies in different stellar mass bins (different panels). In each panel, the contours contain 20, 40, 60, and 80 per cent of the galaxies, while the greyscale shows the overall galaxy distribution. This figure shows that rotational support becomes weakly correlated with angular momentum in massive galaxies.}
	\label{fig:compare_kappa_jstar}
\end{figure*}

Fig. \ref{fig:compare_kappa_jstar} shows that the correlation between $\kappa_{\rm rot}$ and $j_{\ast}$ is stronger at low masses (left-hand panels), becoming weaker in more massive galaxies (right-hand panels). At very high masses, $M_{\ast} \gtrsim 10^{11.5} \Msun$, the galaxy sample is composed of dispersion-dominated spheroids with low $\kappa_{\rm rot}$ values, but which nevertheless exhibit a wide range of specific angular momenta. This suggests that, in general, galaxy mergers might not decrease the angular momentum of the stellar component, but merely redistribute it to larger radii. This inside-out transport of angular momentum had already been observed for DM \citep{Zavala2008, Sharma2012}.

\begin{figure*}
  \centering
	\includegraphics[width=17.5cm]{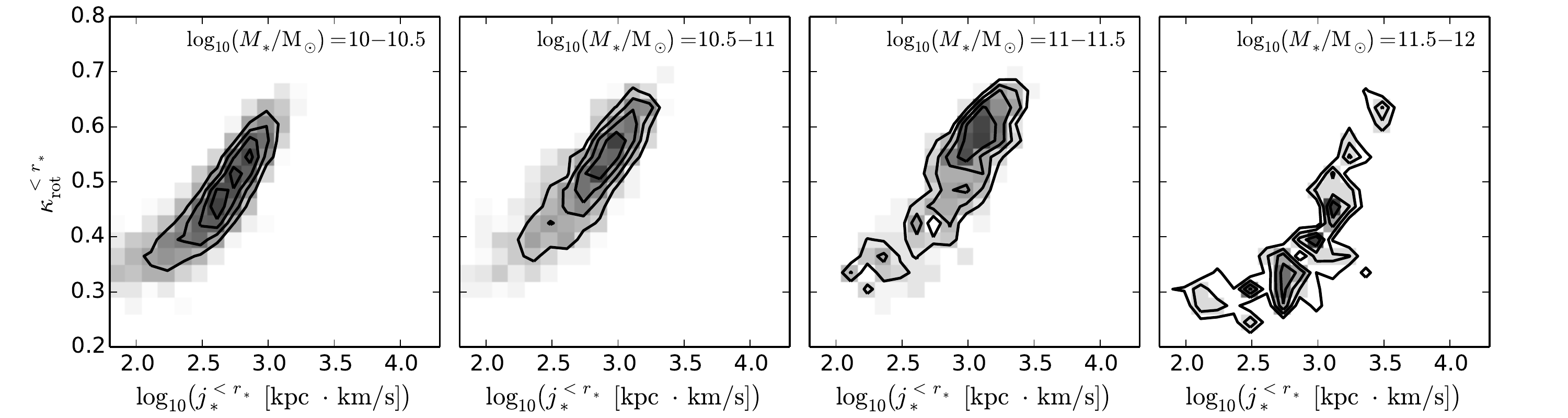}
	\caption{Same as Fig. \ref{fig:compare_kappa_jstar}, but measuring $\kappa_{\rm rot}$ and $j_{\ast}$ within a fiducial aperture corresponding to twice the stellar half-mass radius, $r_{\ast} \equiv 2 \, r_{\rm half, \ast}$. In order to have a fair comparison with Fig. \ref{fig:compare_kappa_jstar}, the galaxies are also binned according to their total stellar mass $M_{\ast}$. This figure shows that $\kappa_{\rm rot}$ and $j_{\ast}$ are always reasonably well correlated if measured within the central region of the galaxy that contains most of the light and stellar mass. However, this comes at the cost of excluding most of the angular momentum in massive galaxies.}
	\label{fig:compare_kappa_in_rad_jstar_in_rad}
\end{figure*}

Fig. \ref{fig:compare_kappa_in_rad_jstar_in_rad} is similar to Fig. \ref{fig:compare_kappa_jstar}, except that $\kappa_{\rm rot}$ and $j_{\ast}$ have been calculated by considering only stars within two stellar half-mass radii, a region which, for practical purposes, is often considered to be `the galaxy' in many simulation studies. Within such an inner region, $\kappa_{\rm rot}$ and $j_{\ast}$ appear to be well correlated even in the most massive galaxies. Nevertheless, the specific angular momentum of massive, dispersion-dominated (low $\kappa_{\rm rot}$ values) galaxies appears to have decreased by an order of magnitude, suggesting that a large fraction of their angular momentum must be found beyond two stellar half-mass radii.

\begin{figure*}
  \centering
	\includegraphics[width=17.5cm]{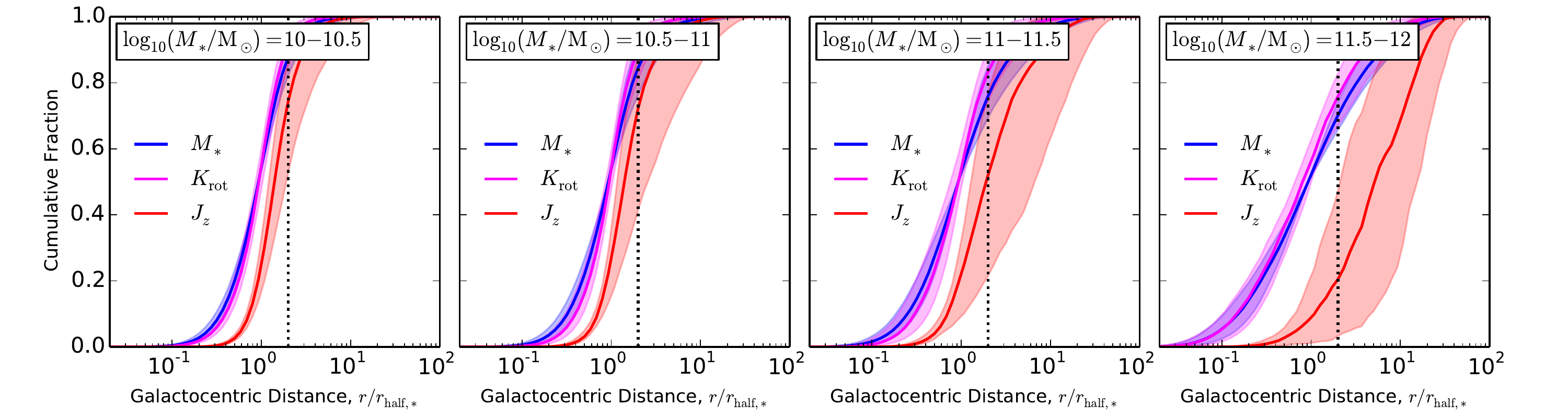}
	\caption{Cumulative radial profiles showing stellar mass (blue), kinetic energy contained in the azimuthal component of the stellar velocities (magenta), and angular momentum along the $z$-axis (red), calculated for galaxies in different stellar mass bins (different panels). Each line shows the median trend for galaxies in the corresponding stellar mass bin, while the shaded regions indicate the 16th to 84th percentile ranges, or 1$\sigma$. Note that most of the angular momentum in massive galaxies is found beyond two stellar half mass radii, i.e. outside of the region that is usually regarded as `the galaxy.'}
	\label{fig:kappa_jstar_profiles}
\end{figure*}

In order to show such radial trends more clearly, Fig. \ref{fig:kappa_jstar_profiles} shows cumulative radial profiles of stellar mass $M_{\ast}$ (blue), $K_{\rm rot}$ (magenta, as defined in equation \ref{eq:kappa}), and angular momentum along the $z$-axis (red), considering galaxies in the same stellar mass bins as the previous two figures. The solid lines and shaded regions correspond to the median trends and 1$\sigma$ ranges, respectively. In each panel, the vertical dotted line is located at two stellar half-mass radii. This figure illustrates perhaps the most important difference between $\kappa_{\rm rot}$ and $j_{\ast}$: the former quantity is essentially `mass-weighted,' as seen from the fact that the blue and magenta lines follow each other quite closely, whereas angular momentum is to some extent `distance-weighted,' such that a few stars orbiting at large radii can have a substantial impact on the galaxy's angular momentum.

\end{document}